\documentclass{aa}  
\usepackage{graphicx}
\usepackage{figsize}
\usepackage{txfonts}
\usepackage{multicol}
\usepackage[usenames]{color}
\usepackage[breaklinks, colorlinks, citecolor=blue, linkcolor=blue]{hyperref}
\usepackage{comment}
\usepackage{longtable}

 \usepackage{tablefootnote}
  \usepackage{comment}
  \usepackage{soul}
\usepackage{ulem}

\begin{document} 

   \title{The evolution of lithium in FGK dwarf stars}
 \subtitle{Influence of planets and Galactic migration}

   \authorrunning{Llorente de Andr\'es et al.}
   \author{F. Llorente de Andr\'es
    \inst{1,2}\fnmsep\thanks{fllorente@cab.inta-csic.es},
     R. de la Reza\inst{3},
    P. Cruz\inst{1},
    D. Cuenda-Muñoz \inst{1},
    E. J. Alfaro \inst{4},
    C. Chavero\inst{5,6},
    C. Cifuentes\inst{1}
    }

    \institute{ 
    Centro de Astrobiolog\'{\i}a (CAB), CSIC-INTA, Camino Bajo del Castillo s/n, campus ESAC, 28692, Villanueva de la Ca\~nada, Madrid, Spain
    \and
    Ateneo de Almagro, Secci\'on de Ciencia y Tecnolog\'ia, 13270 Almagro, Spain
    \and
    Observatório Nacional,  Rua General Jos\'e Cristino 77, 28921-400 São Cristovão, Rio de Janeiro, RJ, Brazil
    \and
    Instituto de Astrofísica de Andalucía (CSIC),Glorieta de la Astronomía s/n, E-18 008 Granada, Spain
    \and
    Observatorio Astron\'omico de C\'ordoba, Universidad Nacional de C\'ordoba, Laprida 854, 5000 C\'ordoba, CONICET, Argentina
    \and
    Consejo Nacional de Investigaciones Cient\'ificas y T\'ecnicas (CONICET), Godoy Cruz 2290, Ciudad Aut\'onoma de Buenos Aires, Argentina
 }
  
\date{Accepted 02 January 2024}

\abstract{

This work aims to investigate the behaviour of the lithium abundance in stars with and without detected planets. Our study is based on a sample of 1332 FGK main-sequence stars with measured lithium abundances, for 257 of which planets were detected. Our method reviews the sample statistics and is addressed specifically to the influence of tides and orbital decay, with special attention to planets on close orbits, whose stellar rotational velocity is higher than the orbital period of the planet. In this case, tidal effects are much more pronounced.
The analysis also covers the orbital decay on a short timescale, with planets spiralling into their parent star. Furthermore, the sample allows us to study the relation between the presence of planets and the physical properties of their host stars, such as the chromospheric activity, metallicity, and lithium abundance. In the case of a strong tidal influence, we cannot infer from any of the studies described that the behaviour of Li differs between stars that host planets and those that do not. 

Our sample includes stars with super-solar metallicity ([Fe/H]>0.15\,dex) and a low lithium abundance (A(Li) <1.0\,dex). This enabled us to analyse scenarios of the origin and existence of these stars. Considering the possible explanation of the F dip, we show that it is not a plausible scenario. Our analysis is based on a kinematic study and concludes that the possible time that elapsed in the travel from their birth places in the central regions of the Galaxy to their current positions in the solar neighbourhood is not enough to explain the high lithium depletion. It is remarkable that those of our high-metallicity low-lithium stars with the greatest eccentricity (e>0.2) are closest to the Galactic centre. A dedicated study of a set of high-metallicity low-Li stars is needed to test the migration-depletion scenario.

}


   \maketitle
%

\section{Introduction}\label{sec:intro}

Different behaviours of the lithium abundance, A(Li), in FGK main-sequence stars with or without planets have been discussed in the literature for the past decades.
For instance, \citet{Gon00} proposed that solar-type stars hosting planets (harbouring mainly giant planets at that time) presented a tendency to have {lower} lithium abundance than similar stars without detected planets. Even if this proposition was soon contradicted by \citet{Rya00}, it transformed into an intense and active debate that has not obtained a clear answer to date.
Several studies \citep[e.g.][]{Isr04,Tak05,Che06,Cas09,Tak10,Gon10,Del14,Sou16,Agu18} suggested that the lithium depletion in stars with planets (hereafter, Y stars) is different from that of similar stars without detected planets (hereafter,  N stars). 
However, other works \citep[e.g.][]{Rya00,Luc06,Bau10,Ghe10,Ram12,Ben18,Cha19} argued that this effect cannot be distinguished.

In general, the aforementioned works primarily focused on presenting and discussing the statistics of Y and N stars (those with or without planets).
The physical causes for the mentioned connection were described by all authors as the presence or absence of generally involved parameters and their effects, such as the stellar metallicity, mass, and age. 
In general, there is an absence of a detailed numerical discussion of a physical model that would trigger this star-planet connection.  
Independently of the two groups Y and N, a detailed model that was generally applied to the proposal by \citet{Gon00} was published by \citet{The12}.
They proposed that the fall of planetary material during the early stages of protoplanetary evolution could modify the stellar lithium abundance. 
When a host star interacts with its planets, in particular, when these planets are massive and on close orbits, it can cause a merging effect during the main-sequence evolution that leads to a range of different effects \citep[see e.g.][who describe the effect of stellar and planetary factors on the orbital evolution of hot Jupiters]{Laz21}.

Recently, detailed models of stars shed new light on the process of lithium enrichment produced by the fall of planetary bodies. \citet{Ste20} confirmed that exoplanets have been observed around stars in several stages of stellar evolution. In many cases, they orbit in configurations that will eventually cause the planets to be engulfed or consumed by their host stars, such as hot Jupiters or ultra-short period planets. \citet{Soa21} proposed that planetary engulfment events are the mechanism of lithium enrichment and contribute to the population of Li-rich giants, with A(Li) $\geq$ 1.5\,dex. More recently, \citet{Sev22} developed models for the lithium enrichment in solar-mass dwarf stars.

In addition to the above, we also specifically address the influence of tides and orbital decay, with special attention to planets on close orbits, whose the stellar rotation velocity is higher than the orbital period of the planet. 
In these conditions, the tidal effects are much stronger. 
Our analysis also covers the orbital decay, associated with the time of the planet spiralling onto its parent star. 
Additionally, we aim to investigate the relation between the presence of planets and the host-star properties, such as chromospheric activity,  metallicity, and lithium abundance.

Our sample includes a few high-metallicity and low-Li stars whose presence in our environment requires an explanation. Previous studies indicated that the scenario designed by \citet{Gui16,Gui19} explains the presence of these old stars in the solar vicinity, suggesting they originated in the inner part of the Galaxy and migrated to their current position. Other studies have proposed alternative scenarios. \citet{Cha21} posited that the original Li abundance in metal-rich Galactic environments might be even higher than the mean value derived in their work, and they suggested that Li depletion might be due to atomic diffusion. In a more recent work, \citet{Dan22} argued that stars with high metallicity and a low lithium abundance migrated from the inner part of the Galaxy, in agreement with the scenario proposed by \citet{Gui19}. They suggested that two mechanisms acted together, stellar evolution and radial migration. They reached this conclusion, but only tentatively, based on chemical abundance research. 

Based on the minimum distance of the star to the inner part of the Galaxy, we discuss here whether stellar radial migration together with the internal process is a plausible scenario to explain the presence of super-solar metallicity star with low A(Li) in the solar vicinity.

This work is organised as follows: In Sect.~\ref{sec:sample} we describe the origin of our stellar sample. 
Section \ref{sec:interaction} is devoted to the interaction of the planet and star, including cases where planets are in close-in orbits. 
In Sect.~\ref{sec:influence} we analyse physical effects and their influence on the lithium depletion. 
In Sect.~\ref{sec:galactic} we discuss the Galactic origin of several metal-rich stars with low lithium abundances in our sample of stars. 
In Sect.~\ref{conclusions} we finally present our conclusions.

\section{Stellar sample}\label{sec:sample}

A full description of our sample was given in our previous works\citep[][hereafter, CH19, LA21, and RF21, respectively]{Cha19,Llo21,Roc21}. 
Table~\ref{t:parameters0} is complementary to the table published in our previous catalogue (see LA21). It lists the planetary systems we studied , together with information on the number of planets in the system, the total planetary mass, their shortest orbital periods, the smallest semi-major axis, and the lowest eccentricity. All of these data were collected from The Extra-solar Planets Encyclopaedia\footnote{Available online at \url{http://exoplanet.eu}, version of September 2022.}. We also include the calculated orbital decay (OD), as described in detail in Sect.~\ref{sec:interaction}. 

We refer to the stars with planets as Y stars and to stars with non-detected exoplanets, as N stars. Our sample contains 1075 N stars and 257 Y stars with known lithium abundances, which have been determined homogeneously as described in Ch19, LA21, and RF21. 
The sample has a large number of planet-hosting dwarf stars, with spectral types from F to K, regardless of the adopted technique used to detect their companion planets, with a total of 330 planets in the sample. We compare our sample with that of other authors who also studied the influence of planets on the behaviour of lithium \citep{Gui16,Agu18,Ben18}. 
It is relevant to mention that the stars in \citet{Agu18} were collected from different sources (see the references therein) and were well characterised afterwards. 
Our work points to thin-disc stars, while \citet{Agu18} included halo K-type dwarf stars. We also reviewed the AMBRE catalogue used by \citet{Gui16}. The most relevant difference between the two samples is that \citet{Gui16} cut off projected rotation velocities higher than 10 and 15\,km\,s$^{-1}$ to derive A(Li) values, while there are no restrictions in our catalogue. 
We also considered the catalogue published by \citet{Ben18}, which is similar to ours but includes stars with ages greater than 10\,Ga. After we studied the A(Li) values and the A(Li)-[Fe/H] relation in all four catalogues, we conclude that our sample shows similar distributions, with minor differences in the dispersion of the A(Li)-[Fe/H] relation.

Finally, the percentage of Y stars in our sample with respect to the total is $\sim$24\,\% of the total. This percentage is common in the literature \citep[see for instance][]{Lin03}. 
We excluded brown dwarfs from our analysis of the planet-star interaction to maintain the limit for planetary mass at 13 M$_{\rm Jup}$ at most \citep{Spi11}. 

\section{Lithium-depletion scenarios}\label{sec:interaction}

\subsection{Lithium depletion in stars with and without planets}

\begin{table*}[!ht]
\caption{Average values of A(Li) of N and Y stars as a function of spectral type.}

\label{table1}
\centering
\scalebox{0.9}{
\begin{tabular}{l|cccc|cccc}
    \hline\hline 
    \noalign{\smallskip} 
   & \multicolumn{4}{c|}{N stars}  & \multicolumn{4}{c}{Y stars}\\SpType & A(Li)$\geq$1.5 & n   & A(Li) <1.5 & n &  A(Li)$\geq$1.5 & n   & <1.5    & n\\
\noalign{\smallskip}
\hline 
\noalign{\smallskip}
 F & 2.46$\pm0.07$ & 148 & 1.10$\pm0.10$ & 19 & 2.50$\pm0.06$ & 41 & 0.91 $\pm0.10$ & 8\\
G & 2.24$\pm0.07$ & 405 &	0.71$\pm0.11$ & 396 & 2.26$\pm0.04$ & 89 & 0.75$\pm0.10$ & 79\\	
K & 2.24$\pm0.10$$^\dagger$  & 3 & 0.30$\pm0.11$ &	110 & 2.64$\pm0.07$$^\dagger$ & 1 &	0.41$\pm0.13$ & 33\\
 \noalign{\smallskip}
\hline 
 \noalign{\smallskip}
All stars	& 2.30$\pm0.07$ & 556 & 0.64$\pm0.10$ & 525 & 2.35$\pm0.05$	 & 131 & 0.63$\pm0.10$ & 120 \\
 \noalign{\smallskip}
\hline 
 \noalign{\smallskip}
Average A(Li) & \multicolumn{4}{c}{1.50$\pm0.08$} & \multicolumn{4}{|c}{1.54$\pm0.07$}\\	
 \noalign{\smallskip}
\hline

\end{tabular}
}
\tablefoot{n is the number of stars in a given subsample. $^\dagger$\,These subsets are too small to be representative. }

\end{table*}

\begin{table*}[!ht]
\caption{Average values of A(Li) for the intervals corresponding to the bimodal distribution. }
\label{table2}
\centering
\scalebox{0.9}{
\begin{tabular}{l|cccccc|cccccc}
    \hline\hline 
    \noalign{\smallskip} 
   & \multicolumn{6}{c|}{N stars}  & \multicolumn{6}{c}{Y stars}\\
    SpType & A(Li)<1.3 & n  & 1.3$\leq$A(Li)$\leq$1.7 & n &  A(Li)>1.7 & n  & A(Li)<1.3 & n  & 1.3$\leq$A(Li)$\leq$1.7 & n &  A(Li)>1.7  & n \\
     \noalign{\smallskip}
\hline 
  \noalign{\smallskip}
 F & 0.94$\pm$0.01 & 10 & 1.53$\pm$0.09 & 9 & 2.50$\pm$0.07 & 138 & 0.64 & 3$^\dagger$ & 1.47 & 3$^\dagger$ & 2.55$\pm$0.05 & 38 \\
G & 0.67$\pm$0.12 & 342 & 1.30$\pm$0.08 & 90 & 2.33$\pm$0.07 & 348 & 0.69$\pm$0.10 & 73 & 1.49$\pm$0.06 & 12 & 2.31$\pm$0.05 & 78 \\	
K & 0.15 & 2$^\dagger$ & 1.43 & 2$^\dagger$ & 2.05 & 2$^\dagger$ & 0.39$\pm$0.12 & 29 & \ldots & \ldots & 2.64 & 1$^\dagger$ \\	
 \noalign{\smallskip}
\hline 
 \noalign{\smallskip}
Sample	& 0.51$\pm$0.12 & 476 & 1.51$\pm$0.08 & 101 & 2.38$\pm$0.07 & 489 & 0.59$\pm$0.11 & 107 & 1.49$\pm$0.06 & 15 & 2.39$\pm$0.05 & 121 \\
 \noalign{\smallskip}
\hline
\end{tabular}
}
\tablefoot{n is the number of stars in a given subsample. $^\dagger$\,These subsets are too small to be representative. 
}
\end{table*}

As mentioned above, several authors claimed to have found a correlation between the lithium depletion and the presence of planets \citep[e.g.][among others]{Isr04,Che06,Tak10,Gon14,Sou16}. These authors proposed that stars with planets are more strongly depleted in lithium than others. 
To support their conclusions, they tried to explain this correlation through the slowdown of rotational velocity on the host stars induced by the presence of planets. 
This retention on the stellar rotation strengthens the mixing in the external layers of the host stars and therefore increases the lithium depletion. 
Other authors \citep[e.g.][among others]{Rya00,Luc06,Ghe10,Ben18,Cha19} found a clear correlation between lithium content and stellar age, but no correlation with the existence of known planets. 
It is important to emphasise that our results are based on a large sample of Y stars.

Based on our sample, we found that the differences in the mean values of A(Li) for both complete sets of Y and N stars are not statistically relevant. 
To show this, we divided our sample by spectral types of each category, N stars and Y stars, and by two intervals of the lithium abundances, A(Li)$\geq$1.5\,dex and A(Li)<1.5\,dex, representing stars not depleted in Li and Li-depleted stars, respectively. 
The results are summarised in Table \ref{table1}, where the values show that the behaviour of Li does not differ between Y and N stars. 
This conclusion is also reached in the relation between A(Li) and age, as shown in RF21 (see their Fig.~6). 
The same finding was obtained in the relation between A(Li) and stellar effective temperature ($T_{\rm eff}$) between Y and N stars in RF21 (see their Fig.~4).
LA21 described that the region centred at A(Li) $\sim$1.5\,dex is poorly populated and RF21 showed that this zone was separated by a bimodal distribution of Li-rich and Li-poor stars. 
The limit of this zone is defined between A(Li) $<$ 1.3\,dex and A(Li) $>$ 1.7\,dex. The authors considered that many of the stars, those with a spectral type F, with A(Li) $<$ 1.3\,dex are evolved Li-dip stars. Then the low number of stars in this area could explain the difference in the A(Li) in the case of F-type stars. 
For G-type stars, the difference in A(Li) could be explained by the difference in the [Fe/H], which is higher for Y stars than for N stars (approximately [Fe/H] of 0.014\,dex against $-$0.117\,dex, respectively).

Performance of identical calculations for the three intervals yields the same conclusion, as indicated by Table \ref{table2}. In other words, the presumed influence depends on the nature of the stars, not on the presence of planets.
These results contradict the authors who stated that stars with planets are more depleted in Li than the others. 
We suspect that this disagreement stems from the difference in sample size.

\subsection{Orbital decay and lithium depletion}
 
The results presented above motivated us to further investigate the lithium-depletion behaviour in other special situations that are fundamentally based on star-planet interactions, such as tidal influence and orbital decay.
These are fundamentally related to the proximity of planets -- until they fall into their host stars or are engulfed by them.

Tidal dissipation in planet-hosting stars could play an important role in the evolution and survival of hot-Jupiter systems \citep[e.g.][]{Lec10, Mat10}. Moreover, according to \citet{Ham19}, the stellar population that hosts hot-Jupiter planets is a younger subgroup of field stars according to their lower Galactic velocities. This argument is accompanied by the fact that planets are destructed by tidal effects during the whole main-sequence phase \citep[see also][]{Bar20}. Definitively, star-planet interactions could result in a modification of the structure and behaviour of the convective layer, which in turn would result in a stellar lithium-depletion behaviour.

Tidal forces result from the varying gravitational fields over one or both bodies. The fields deform their shapes, which may be further modified by rotation. This deformation results in energy dissipation, which then evolves irreversibly, in size and shape. 
Tides are generated both on the planet through the gravitational potential of the star and on the star through the planet. 
Tidal forces depend on the stellar rotation compared with the orbital period of the planet, which typically results in a decrease in the orbital eccentricity, connected with an orbital decay \citep[see e.g.][]{Ogi14}. This means that OD is one of the processes that is considered to deliver planets to close-in orbits, which is a stellar tidal effect on the planet when they become strong. This justifies our following subsection \ref{sec34}, which is devoted to studying the influence of planets when they orbit faster than their parent star rotates \citep[see][]{Per18}. 
As the planet approaches, tidal forces would play an important role on the star, which may influence the Li depletion, as we describe in the following paragraphs. 
Our goal is to explore whether the interaction between stars and planets may increase the number of mixing layers deep into the stellar convective zone, and may increase them enough for Li to be burned and therefore depleted.

\citet{Bar09,Bar11} assumed that tides rise on the central stars by planets on evolving circular orbits. The majority of planets with short orbital periods are observed to have a low eccentricity, which supports tidal circularisation theory. This means that the eccentricity of the orbit is reduced over time, so that the orbit becomes increasingly less elliptical and many times closer, which favours the spiral movement of the planet towards the star \citep{Sun08}.

As the planet approaches its parent star, the mutual influence becomes more pronounced. Because orbital decay is the process considered to deliver planets to close-in orbits and hence increases their mutual influence, we studied this influence based on OD values. These values were computed using Eq.\,\ref{eq1}, which is quoted from \citet[][see references therein]{Lai12}, 

\begin{equation}
    {\rm OD} \approx 1.28\, \frac{Q'}{10^7}\, \left(\frac{\rho_\star}{\rho_{\odot}}\right)^{5/3} \left(\frac{M_\star}{10^3\,M_p}\right)\, P^{13/3} \,\, {\rm (Ga)},
    \label{eq1}
\end{equation}

\noindent
where $Q'$ is the tidal quality factor, which is the measure of orbital energy dissipated per orbit, due to tidal forces. The $Q'$ value is a number associated with an orbiting body that characterises the rate at which the tidal force converts kinetic energy into heat within such a body. 
$\rho_\star/\rho_{\rm \odot}$ is the density of the star in units of solar density (this ratio varies in our sample from 0.15 to 2.3). $M_p$ is the mass of planets in M$_{\rm Jup}$ (in the case of a multiple-planet system, $M_p$ is represented by the sum of masses of planets, $\Sigma M_p$). $M_{\star}$ is the mass of the stars in units of M$_{\rm \odot}$. Lastly, $P$ is the orbital period, in days. $OD$ is given in Ga.

We adopted $Q'$ = 10$^{7}$ (or $\log{Q'}$ = 7), following several authors \citep[e.g.][]{Pen18,Col18,Ham19}. Our estimated OD values are presented in Table \ref{t:parameters0}. For multi-planet systems, we conjectured that OD represents the value of OD of a planet that would concentrate all the planetary mass into a single orbit formed by the shortest period, the lowest separation, and the lowest eccentricity with which the value of OD would be lowest. This implies a shorter time to decay, and therefore, the greatest star-planet influence scenario.

We also explored other tidal definitions such as the tidal plunging time, $\tau_{\rm plunge}$, as described by \citet{Met12} in their Eq.~1. They conducted several simulations with different values of $Q'$ and concluded that it is viable if $Q'$ = 10$^{6}$, but unlikely if $Q'$= 10$^{8}$ or greater, which again supports our choice of $Q'$= 10$^{7}$. Although we are aware that the equation by \citet{Met12} is usually applied in a different context, we estimated $\tau_{\rm plunge}$. We also explored the OD time as defined by \citet[][Eq. 58]{Bar20}, also being aware that it is applied in another context beyond our scope. The relations between these two equations and the equation adopted by us, given by \citet{Lai12}, are shown to be linear:

\begin{equation}\label{eq2}
\begin{split}
 \log\tau_{\rm plunge}\, {\rm (Metzger)}\,=\, 0.94\, \log {\rm OD}\,{\rm (Lai)} - 2.09,  R^2 = 0.94;\\
 \log {\rm OD}\, {\rm (Barker)} = 1.64 \log {\rm OD}\, {\rm (Lai)} + 2.53,  R^2 = 0.99.
 \end{split}
\end{equation}

\noindent
This relation between $\tau_{plunge}$ and OD (Lai) can be misleading. If we were to substitute the density of the star for that of a rocky planet, then both values would be the same, but these equations are applied in different contexts. 
In Barker's case, the distinction lies in the different components in the equations, such as separation and density. 
Hence, we continued to adopt Eq. \ref{eq1}.

The fate of a planet as it approaches its parent star also depends on the stellar density, commonly referred to in terms of the density ratio (see Eq.\,\ref{eq1}). \citet{Met12} related the approach of the planet, and consequently its engulfment, with the density of the planet in such a way that the higher the density of the planet, the higher the probability that it is submerged in the stellar envelope, which can also destroy the planet. 

The lithium depletion is not affected. To show this, we proceeded as follows. We computed OD, using Eq.\,\ref{eq1}, and limited its value by comparing it to the conclusions reached by \citet{Ma21}. These authors showed that short-period massive planets can be quickly destroyed by non-linear mode damping, while short-period low-mass planets can survive even when they undergo substantial inward tidal migration. After several interactions, we found that their affirmation is accomplished for $\log{\rm OD} \leq -0.3$, that is, OD $\leq$ 0.5\,Ga. Understanding these values allows us to explore the impact on lithium behaviour in such circumstances.

By delimiting the orbital decay to a value of OD $\leq$ 0.5\,Ga, our sample was reduced to 41 stars, 25 of which have $\Sigma M_p$ $\geq$1\,M$_{\rm Jup}$, with a range of periods between approximately 1 and 6 days. These systems have a very narrow separation range, $a$, between 0.02 and 0.07\,au, and their eccentricity, $e$, is low or null, between 0.0 and 0.07, except for HD~42176, HD~118203, HD~125612, and HD~147506, whose eccentricities lie between 0.2 and 0.5. Only 6 stars have A(Li) < 1.5\,dex. Sixteen of the original 41 stars have lower planetary masses ($\Sigma M_p$ < 1\, M$_{\rm Jup}$) and exhibit similar ranges of period, from separations of approximately 2 and 7 days of separation, with $a$ between 0.03 and 0.08\,au, and eccentricities from 0.0 to 0.09, except for HD~185269, whose $e\sim$0.3. Only 4 stars are depleted in Li, that is, they have A(Li) < 1.5\,dex. In conclusion, for $\Sigma M_p$ higher or lower than 1\,M$_{\rm Jup}$, we find no difference in the percentage of stars with lithium depletion. It is approximately 25\,\% for both subsets. This indicates that the number of stars with Li depletion does not increase due for the low values of ODs, which are the planets that orbit more closely to their star, which means a stronger influence of the tides.

In conclusion, when $\log$ OD is lower than or equal to $-$0.3, then the mean separation is $a\sim$0.05\,au and the mean orbital period is shorter than 3.5 days. These conditions are very favourable for a planet to have a high probability, action by the tides, of falling onto the star. Nevertheless, the lithium depletion is not different, even in the circumstances.

\subsection{Engulfment of planets and the lithium-enrichment process}

In the previous section, we explained that under certain conditions of close separation, short orbital periods, and a relatively low value of OD ($\leq$ 0.5\,Ga), there is a high probability that the planet falls onto its star, which may affect the behaviour of Li by accretion. For instance, \citet{Dea15} stated that stars that accrete more than 0.3\,M$_{\rm Jup}$ are expected to be subjected to strong Li depletion.
This motivated us to study the effects of a planet-star interaction in Sun-like stars, considering the effects of planetary orbits until the planet plunges onto the host star, resulting in the destruction of the planet. This process preferentially occurs in stars hosting hot Jupiters and enable us to measure the lithium enrichment in the host stars produced by planets, as described below.

Several works in the literature have been dedicated to the study of planetary engulfment and stellar enrichment, for example, \citet{Oh2018} mentioned a potentially strong ([Fe/H]$\sim$ 0.2\,dex) signature of planetary engulfment in the HD\,240429-30 (Kronos-Krios) system. However, engulfment is not common, but rare and very difficult to detect, as shown recently by \citet{Beh22}. 
All previous research works only studied to lithium enrichment in giant stars, where the process of spiral planets is physically different than for dwarf stars \citep[see for instance][and references there in]{Soa21}. 

The results of the ingestion of planets or cores of giant planets depend on the mass of the falling body and on some stellar properties, such as their mass and the size of their convective zones (CZ). It will also depend on processes such as diffusion and thermohaline mixing, which is a hydrodynamic instability caused by differences in composition that dominates during the post-engulfment first $\sim$ 5-45\,Ma, weakening signatures by a factor of $\sim$2 before giving way to depletion via gravitational settling on longer timescales \citep{Beh23}. The combination of all these factors would furnish the signatures of the level of lithium weakening or enrichment, which depends on the lifetimes of the ingested lithium.

Lithium enrichment in solar-mass dwarf stars has recently been performed by \citet{Sev22}. 
Their research was based on MESA stellar models \citep{Pax11,Pax19}, which take into consideration all the physical ingredients for the case of a planet engulfment during the main-sequence stage. They considered a solar composition for the stars and masses ranging between 0.7 and 1.4\,M$_{\rm \odot}$. 
The engulfed bodies had an Earth-like composition, with masses of 1, 10, and 100\,M$_{\rm \oplus}$. In the following paragraphs, we compare the lithium abundance A(Li) when the planetary OD reaches the values discussed in the previous subsection, with the expected long lithium lifetimes resulting from the models by \citet{Sev22}. 
We are aware that these OD values represent only one possible engulfment process. For this purpose, we selected 35 stars hosting hot Jupiters as best candidates because \citet{Sev22} also based their models on hot Jupiters (HJ). We adopted the definition of HJ from \citet{Wri12}. Hot Jupiters are planets with an orbital period of $P\leq10$\,days and $M_p \geq 0.1$\,M$_{\rm Jup}$, so that these 35 stars are single-planet systems whose planet is a HJ. At the same time, the selected objects had to exhibit OD values below 0.5\,Ga, as previously described. These values, in effect, would be those of planetary systems that best represent targets to potentially spiral onto the host star. All these objects are enclosed in Table \ref{t:parameters0}.

\begin{figure}
    \centering
    \includegraphics[width=.95 \columnwidth]{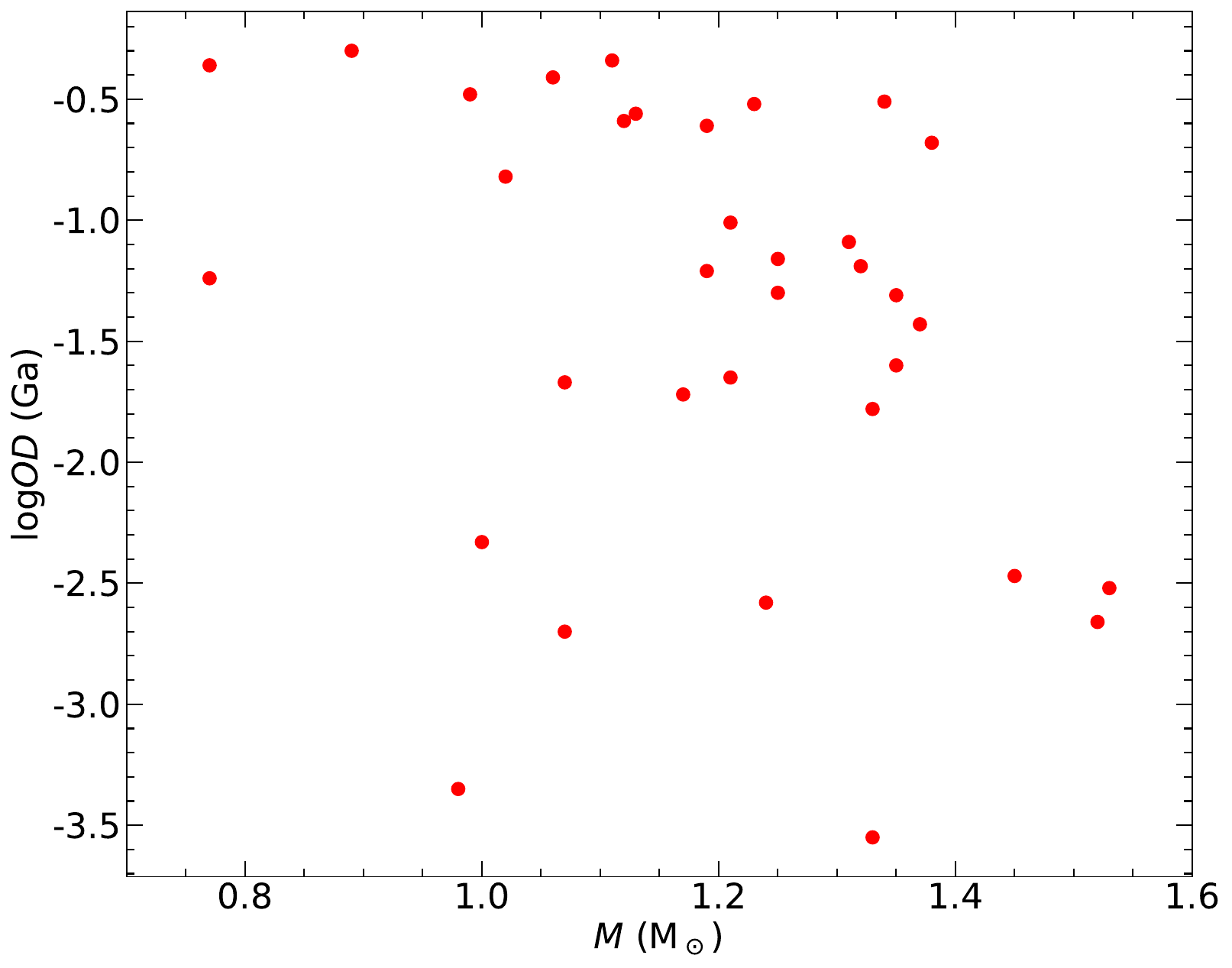}
    \caption{Orbital decay in respect to the stellar mass of hot-Jupiter host stars, whose estimated $\log {\rm OD} <-0.3$ (OD $<$ 0.5\,Ga).
    }
     \label{figure1}
\end{figure}

The $\log$ OD values appear to be well distributed in Fig.~\ref{figure1} for all stellar masses from 0.77 to 1.53\,M$_{\rm \odot}$. This distribution enables us to compare our values of A(Li), when the $\log$ OD values are lower than or equal to $-0.3$, with the signatures of the presence of new lithium enrichment predicted by the models of \citet{Sev22}. These authors mentioned that for masses below 0.7\,M$_{\rm \odot}$, the injected lithium is short lived because the CZ reaches the lithium-burning zone, where lithium is rapidly destroyed. For the mass interval between 0.7 and 1.0\,M$_{\rm \odot}$, the new lithium lifetimes are quite long, 1\, Ga or more. In this last case, the models are independent of the masses of the merging planets, and we derived that the new lithium abundances can reach peaks of A(Li) $\sim$2.0\,dex for a 0.8\,M$_{\rm \odot}$ star and $\sim$2.5\,dex for 1.0\,M$_{\rm \odot}$ star (see Fig.~8 in \citealt{Sev22}).

\begin{figure}
    \centering
    \includegraphics[width=0.99 \columnwidth]{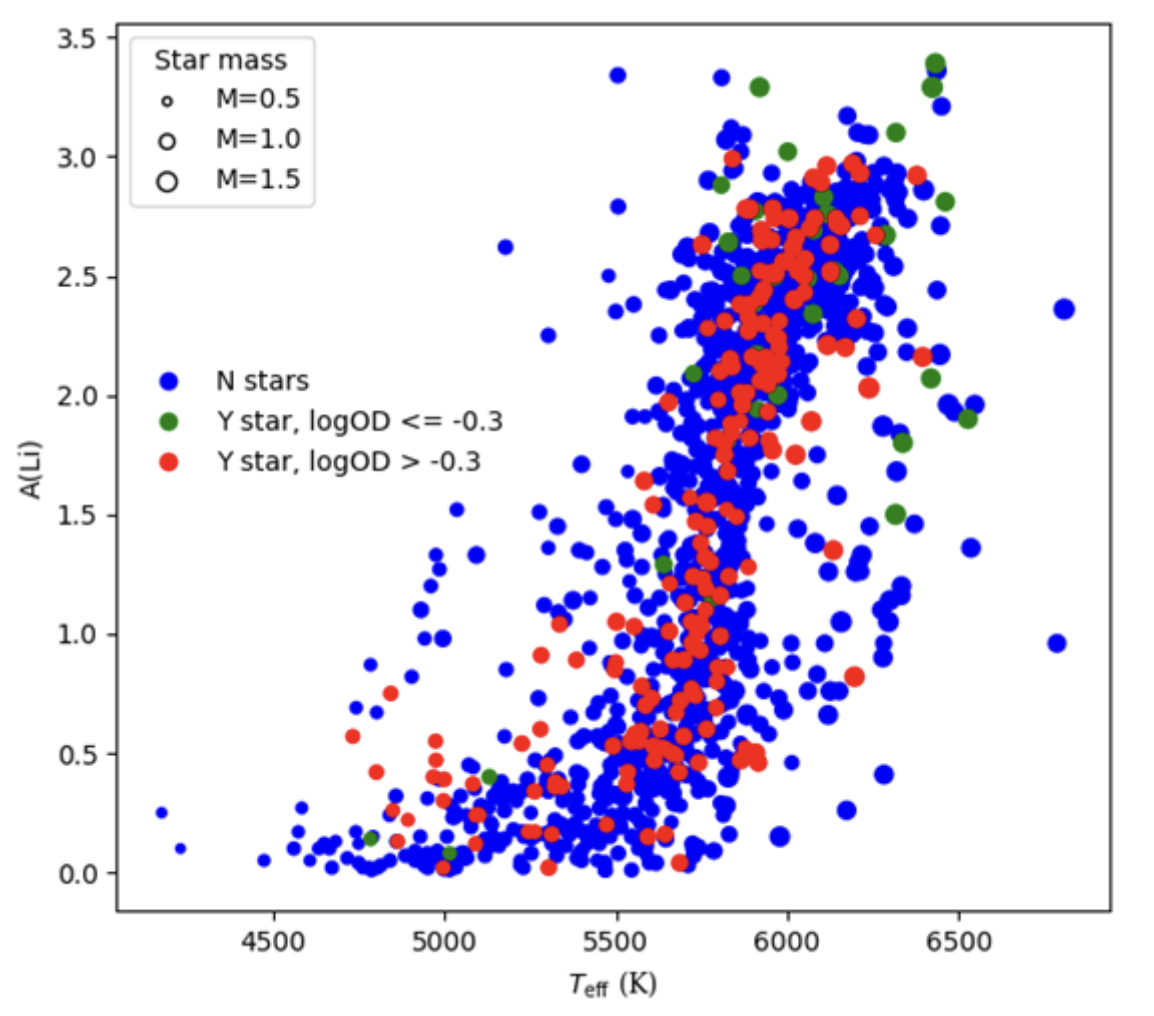}
    \caption{Relation between lithium abundance and $T_{\rm eff}$. N stars are represented by blue dots and Y stars by red dots, except for stars whose $\log{OD} \le -0.3$ are identified as green dots. The symbol sizes are proportional to the stellar masses, as shown in legend of the top left panel.}
     \label{figure2}
\end{figure}

Figure \ref{figure2} depicts the distribution of lithium abundances in Y stars with $\log{\rm OD} \leq -0.3$ as a function of stellar effective temperature ($T_{\rm eff}$) for different stellar masses (illustrated as different symbol sizes). No significant difference is observed between the two sets (N and Y stars); none of the Y stars exhibit larger lithium abundances than N stars within a stellar mass range of 0.77 to 1.5\,M$_\odot$. Additionally, the lithium abundance in Y stars with a stellar mass of approximately 0.75\,M$_\odot$ reaches a lower limit, approaching null A(Li). This location aligns with the nearly zero lower limit shown in \citet{Sev22} (see their Fig.\,8). The same trend is observed for Y stars with stellar masses between 0.9 and 1.0\,M$_\odot$. They are situated at an A(Li) level of approximately 0.5\,dex in figure \ref{figure2}. Consequently, we can conclude that there is no evidence of lithium enrichment, even in the most favourable intervals for lithium signatures with extended new lithium lifetimes, as deduced in the models of \citet{Sev22}.

\subsection{Lithium depletion in the case of planets on close orbits}\label{sec34}

The previously mentioned results led us to impose more critical restrictions with the objective to determine the influence of tides with a great effect on the star-planet system. To evaluate this influence, we further required a small orbital separation ($a\leq$ 0.2\,au) and an orbital period shorter than or equal to the stellar rotation period (P$_{\rm orb}$ $\leq$ P$_{\rm rot}$). Under these conditions, tidal effects are intensified. After these criteria were applied, the number of stars in our Y stellar sample was reduced to 34 objects. 
 
The aim of this subsection is devoted to verify that, even in the strongest mutual star-planet influence, there is no difference between the behaviour of lithium in stars with and without planets. 
Pursue this goal, we created a homogeneous set for the two categories (N and Y stars). We therefore took N stars with the same range of stellar rotation periods, as shown in Table\,\ref{table3}. Under this condition, both subsets can be compared. 
Figure\,\ref{figure2} supports the fact that there is no difference in the behaviour of lithium in stars with planets and in stars without planets.

\begin{table*}[!ht]
\caption{Average values of the stellar parameters for N stars and Y stars, with the same range of stellar rotation periods (P$_{\rm rot}$ in days) and within the two intervals of the abundance of Li: A(Li) $\geq$1.5 and A(Li) < 1.5.} 
\label{table3}
\centering
\begin{tabular}{l|cc|cc}
    \hline\hline 
    \noalign{\smallskip}
& \multicolumn{2}{|c|}{Y stars} &\multicolumn{2}{c}{N stars} \\ 
    \noalign{\smallskip} 
\hline
    \noalign{\smallskip}
    P$_{\rm rot}$ [d] & \multicolumn{2}{|c|}{7.0 -- 48.1} & \multicolumn{2}{c}{7.0 -- 48.0} \\
    A(Li) & $\geq$1.5  &  $<$1.5  &             $\geq$1.5  &  $<$1.5 \\
    $\mathcal{M}_\star$ [M$_\odot$]  & 1.19$\pm$0.03  &  0.93$\pm$0.03   &    1.07$\pm$0.03   &  0.95$\pm$0.03 \\
    $\log{R'_{\rm HK}}$  & $-$4.7476$\pm$0.04  &  $-$4.88$\pm$0.03  & $-$4.74$\pm$0.01  & $-$4.8174$\pm$0.03 \\
    $v\,\sin{i}$ [km\,s$^{-1}$] & 4.12  &  1.672   & 4.08$\pm$0.48   &  2.57$\pm$0.53 \\
    {[}Fe/H{]}  & 0.18$\pm$0.04  &  0.09$\pm$0.03   & $-$0.059$\pm$0.03   &  $-$0.089$\pm$0.03 \\
    $T_{\rm eff}$ [K]  & 5955$\pm$73  &  5492$\pm$113   & 5931$\pm$80   &  5572$\pm$100 \\
    Age [Ga]  & 3.45$\pm$1.10  &  4.34$\pm$2.52  & 3.5$\pm$1.64  &  5.02$\pm$2.66 \\
    M$_{\rm Planet}$ [M$_{\rm Jup}$]  & 1.888  &  1.098   &        \ldots  &  \ldots \\
    $a$ [au]  & 0.05673  &  0.065558  & \ldots &  \ldots \\
    \hline

\end{tabular}
\end{table*}


The only significant difference found between the two samples N and Y is in their metallicity. However, there is no relevant difference with stellar mass, a smooth difference in the projected rotational velocity ($v\sin{i}$), and a slight age difference, as Y stars appear to be younger than N stars, which agrees with \citet{Ham19}. 
In summary, we did not find a relevant influence of the presence of planets on the behaviour of lithium, even after imposing extreme conditions in which the influence of the planets should be greater.

\section{Stellar physical characteristics and their influence on the lithium depletion}\label{sec:influence}

After we verified numerically that the presence of planets and the mutual interaction between planets and their parent stars did not influence lithium depletion, we analysed some physical stellar quantities that might influence the behaviour of the Li abundance. These quantities included chromospheric activity, rotational velocity, and metallicity. 
The variation in these magnitudes may be influenced by the presence of planets, which in turn might affect the lithium behaviour. It is worth noting that stellar binarity is not explored here.

\subsection{Effective temperatures and masses}

We found no different behaviour in the relation between A(Li) and $T_{\rm eff}$ related to the presence of planets, as we showed in Fig.~\ref{figure2}. Even when we considered the two extreme cases for $T_{\rm eff} >$ 5700\,K, where A(Li) $<$ 1.5\,dex and A(Li) $\geq$ 1.5\,dex, the presence of planets does not show any difference. 
When we calculated the mean value for A(Li) for the first situation (i.e. A(Li) $<$ 1.5\,dex), we found a value of around 0.92$\pm$0.11\,dex for N stars and around 0.96$\pm$0.10\,dex for Y stars. In the second case (A(Li) $\geq$ 1.5\,dex), the values are 2.33$\pm$0.07\,dex for the N stars and 2.37$\pm$0.05\,dex for the Y stars. 
These values are different from those presented in Table\,\ref{table3} because we considered our complete sample here. This means that, if there is a difference in the behaviour of lithium with respect to $T_{\rm eff}$, considering either N or Y stars, it is statistically insignificant. Furthermore, in what was previously called the Li desert, with 5950 $<T_{\rm eff}<$ 6100\,K; 1.55 $<$ A(Li) $<$ 2.05\,dex \citep{Ram12}, LA21 demonstrated that it is rather due to a statistical distribution fluctuation than to a real physical fact. Even so, however, within the box that we defined to demonstrate that there were stars in the so-called desert, there were N stars and Y stars. In the interval between A(Li) $<$ 1.3 and A(Li) $<$ 1.7\,dex, detailed in RF21, the presence of stars Y and N is appreciated.

The number of Li-depleted stars those with A(Li) $<$ 1.5) strongly depends on the mass of the parent star. It might be concluded that the parent mass has a strong influence as for stars without detected planets. Thus, for both categories, the lower the stellar mass, the smaller the lithium abundance. Figure\,\ref{figure2} illustrates this affirmation.

\subsection{Magnetic activity and rotation}

We briefly discuss here the influence of the chromospheric activity index from the literature and stellar rotation, combined with the presence of planets, on the behaviour of the lithium abundance.

RF21 showed the Galactic bimodal character of lithium, which is clearly shown in the relation A(Li) versus $T_{\rm eff}$ and is also reproduced in Fig.\,\ref{figure2}. This bimodality can be represented by two centroids, defined by a Kolmogorov–Smirnov test, whose description is detailed in that same work. It is straightforward to deduce that a star has a higher probability to be active if it belongs to the A(Li) $\geq$ 1.5 population and otherwise, if it belongs to the A(Li) < 1.5. We conclude that the two A(Li) populations are distinguished not only by the two slopes of lithium depletion, but also by a different magnetic activity.

\citet{Mis2012} concluded that the lithium abundance and the projected rotational velocity are correlated but the lithium-activity correlation is only evident in a restricted effective temperature range. 
We found in the present study that in the interval around the critical stellar rotational velocity of  $v\sin{i}\sim5\pm0.6$\,km\,s$^{-1}$ (described in LA21), the activity index reaches its highest value of $\log{R'_{\rm HK}} \sim -4.4\pm0.2$ both for stars with and without identified planets. 

\subsection{Metallicity} \label{metallicity}

Several studies were dedicated to the role played by the metallicity and the presence of planets, sometimes presenting diverging results \citep[i.e.][]{Pas07, Mun18, Mul18, Roc21}. From our sample, we derived some significant differences between the mean metallicity for the Y stars, 0.09$\pm$0.035\,dex, and the mean metallicity for the N stars, $-$0.095$\pm$0.033\,dex. However, we were unable to establish a relation between A(Li) and [Fe/H] and found not even a trend. The spread of correlation values confounds any interpretation as the apparition of a greater lithium depletion, even though metallicity induces a more complex effect on the processes in the internal layers in which the lithium is burned. 
A group of stars strikingly has a high metallicity and a very small abundance of lithium, the existence of which is a matter of debate. We discuss this in Sect.\,\ref{sec:galactic}.

\subsection{Additional results}

Some collateral results appear in the details of our sample of host stars. We describe them below.

\subsubsection{Metallicity and planetary mass}

\begin{figure}
    \centering
    \includegraphics[width=.99 \hsize]{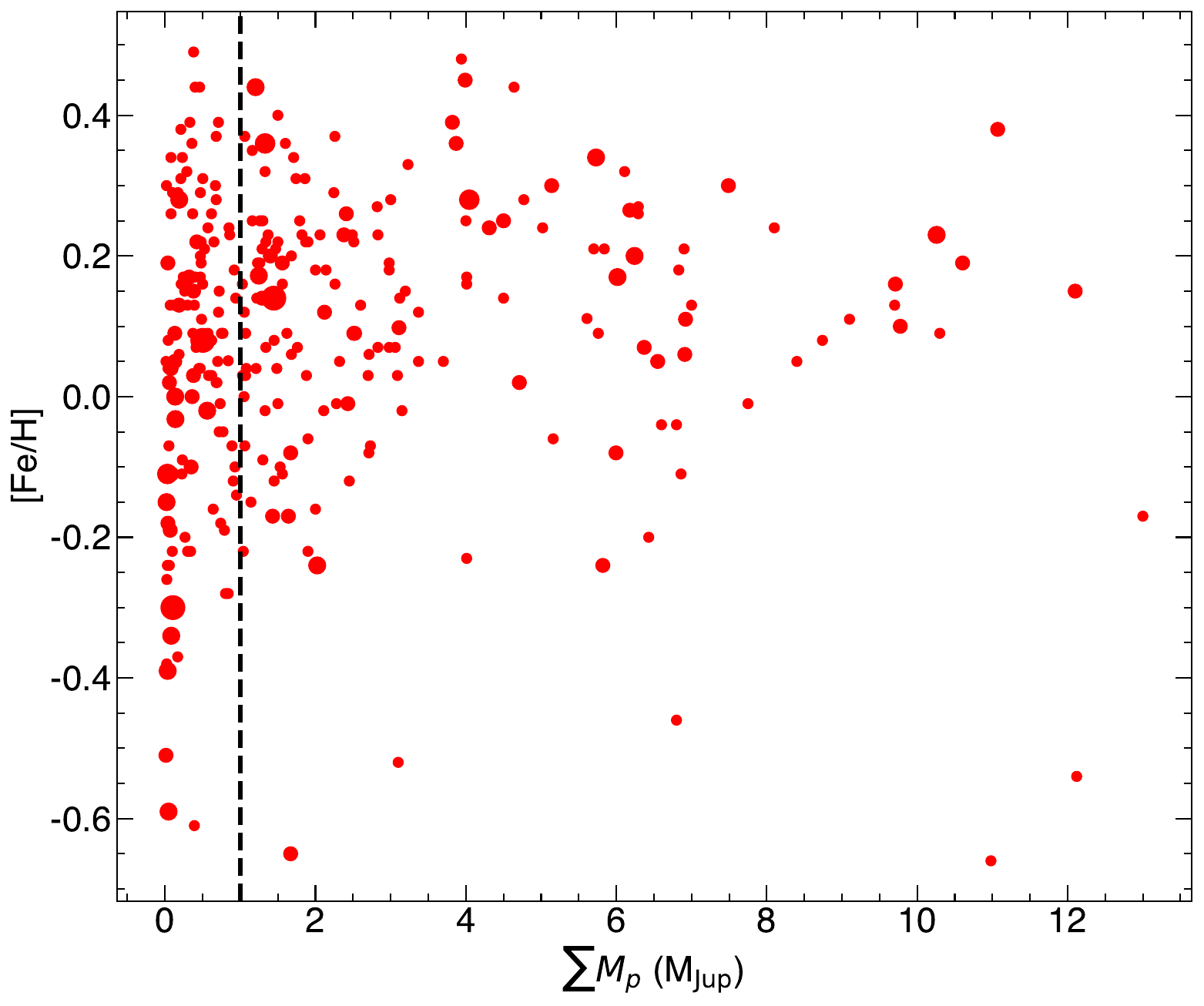}
    \caption{
    Stellar metallicity as a function of total planetary mass in units of M$_{\rm Jup}$. The different circle sizes are proportional to the total number of planets in the system.
    }
     \label{figure3}
\end{figure}

\begin{figure*}
    \centering
    \includegraphics[width=.45 \hsize]{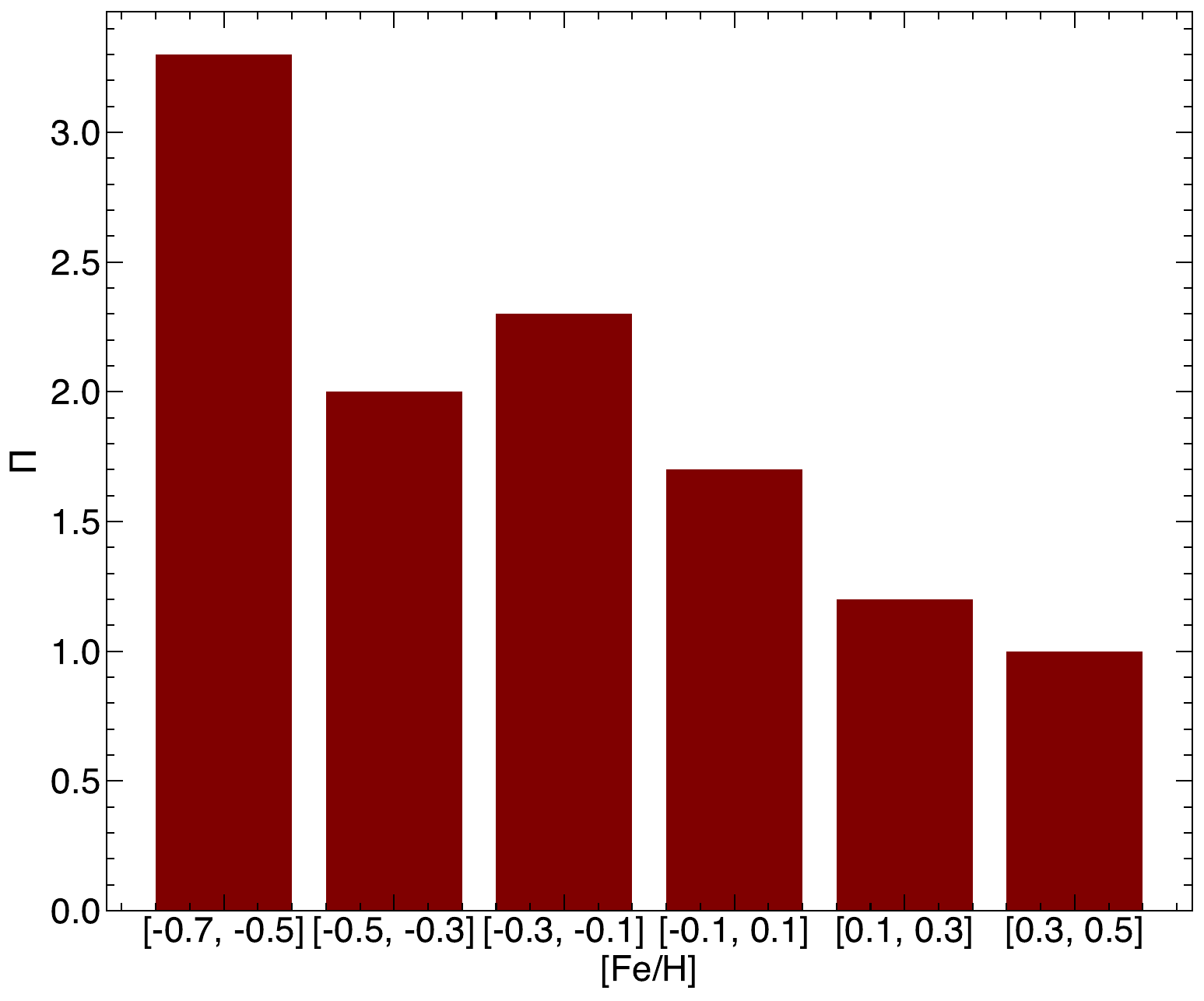}
    \includegraphics[width=.45 \hsize]{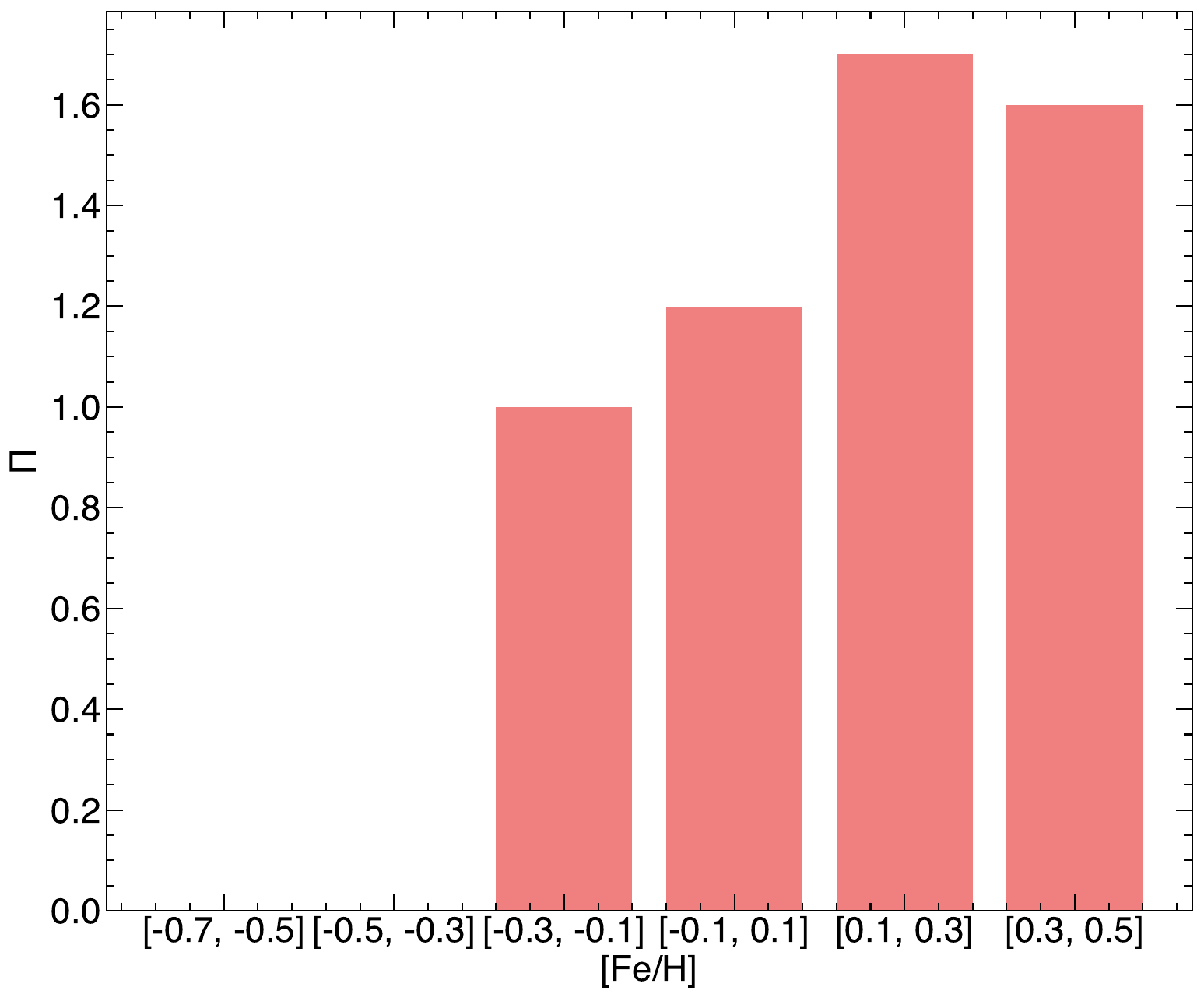}
    \caption{
    Bar plot of $\Pi$, the ratio of the total number of planets and the total number of stars, for different metallicity bins and systems in close-in orbits ($a \leq 0.2$\,au). The left panel shows planetary systems with a low planetary mass ($\Sigma M_p$< 1\,M$_{\rm Jup}$) and the right panel shows those with a higher mass ($\Sigma M_p \geq$ 1\,M$_{\rm Jup}$).
    }
     \label{figure4}
\end{figure*}

It is well known that metallic material is necessary to form planets (e.g. CH19, and references therein), but it has not been yet exhaustively studied that when the iron abundances are low (i.e. in the initial epochs of the Galactic life, when the thick-disc stars were formed at a [Fe/H] < $-$0.2\,dex), other abundant elements, with condensation temperatures similar to that of iron, can replace the insufficient iron to build planets \citep[][and references therein]{Del23}. 
We examined the relation between stellar metallicity, which was taken as the iron-to-hydrogen abundance ratio, and the mass of planets, and the possible correlation with the number of planets that cohabit in the planetary system. Fig.\,\ref{figure3} presents the relation between the stellar metallicity and the total mass of its hosted planets. This shows that the metallicity is less scattered when $\Sigma M_p$ $\geq$ 1.5\,M$_{\rm Jup}$ than when $\Sigma M_p$< 1\,M$_{\rm Jup}$, adding that the dispersion of metallicity values is related to the mass of the parent star. The lower the stellar mass, the greater the dispersion.
This figure also shows that a star with a lower metallicity can form small planets. This allowed us to search in more detail for the behaviour of metallicity in comparison with the total mass of the planets in the systems in our sample, as described below. Moreover, we can evaluate the variation in the ratio (labelled $\Pi$) of the total number of planets and the total number of stars for any subsample defined by common parameters. We focused on the four cases described in Table\,\ref{table4}, where $\Pi$ approximately varies between 1.4 and 1.8. According to the currently available sample, the number of planets present in a planetary system is mainly a function of the host star mass, as it seems that the lower the stellar mass, the greater the number of planets it hosts, regardless of the planetary mass. It is worth emphasising that our sample is not free of observational bias.

\begin{table}[!h]
\caption{Group of stars based on planetary parameters.}
\label{table4}
\centering
\begin{tabular}{lcccc}
    \hline\hline 
    \noalign{\smallskip}
    Case & $a$ & $\Sigma$ $M_p$ & ${M}_\star$ & $\Pi$   \\
     & [au] & [M$_{\rm Jup}$] & [M$_\odot$] &   \\
    \noalign{\smallskip}
    \hline
    \noalign{\smallskip}
    1 & $\leq$ 0.2 & $\leq$ 1 & $\leq$ 1 & 1.80 \\
    2 & $\leq$ 0.2 & $\leq$ 1 & >1 & 1.40 \\
    3 & $\leq$ 0.2 & >1 & $\leq$ 1 & 1.60 \\
    4 & $\leq$ 0.2 & >1 & >1 & 1.45 \\
\noalign{\smallskip}
\hline
\end{tabular}
\end{table}

The analysis of our sample leads us to the conclusion that large planets form in areas of high metallicity when the masses of the stars exceed the solar mass, but small planets are formed even in lower-metallicity environments when the mass of the parent star is lower than one solar mass. It also shows that giant planets form in areas of higher metallicity, regardless of the mass of the parent star. 

However, when we restrict the set of planets to those with close-in orbits ($a\leq$ 0.2\,au), the four cases that are presented in Table\,\ref{table4}, $\Pi$ varies with the metallicity in a way that the lower the metallicity, the higher $\Pi$ when $\Sigma M_p \leq$ 1\,M$_{\rm Jup}$. The opposite but not as drastic case is when $\Sigma M_p$ $>$ 1\,M$_{\rm Jup}$. That holds for multi-planet or single-planet systems. This is illustrated in Fig.\,\ref{figure4}.

\subsubsection{Close-in orbits and short-period planets}

\begin{figure}
    \centering
    \includegraphics[width=.99 \hsize]{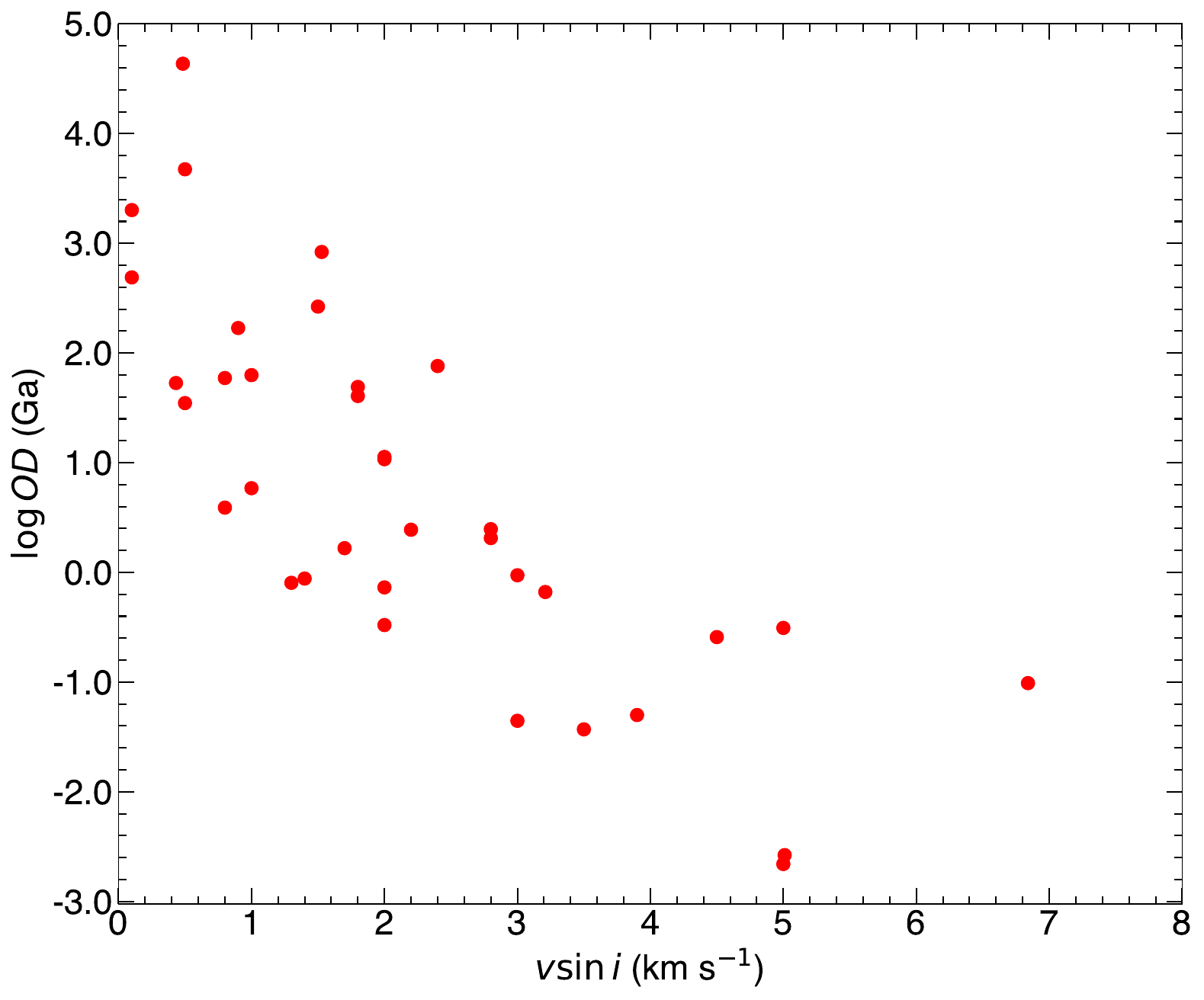}
    \caption{
The logarithm of the orbital decay as a function of $v \sin{i}$ for stars with confirmed planets and restricted to those with P$_{\rm orb} \leq$ P$_{\rm rot}$ and $a <$ 0.2\,au.
    }
     \label{figure5}
\end{figure}

Recalling the restrictions imposed in subsection \ref{sec34}, a Y star with P$_{\rm orb} \leq$ P$_{\rm rot}$ and $a\leq$ 0.2\,au, we plot Fig.\,\ref{figure5}, showing the relation $\log{\rm OD}$ as a function of $v\sin{i}$. This corroborates that as the planet approaches the star, the latter increases its rotation speed. Moreover, sufficiently massive closely orbiting planets are significantly affected by the tidal dissipation of the parent star to a degree that is parameterised by the tidal quality factor $Q'$. This process speeds up the projected rotational velocity of the star, but not over the critical velocity of $v\sin{i}\leq$ 5\,km\,s$^{-1}$, except for HD\,179949, which is a BY Dra variable star, while reducing the planetary semi-major axis and leading to its tidal destruction. In other words, they spin up their stars significantly while spiralling to their deaths. This conclusion can be drawn from Fig.~\ref{figure5}. 
Therefore, the lower the OD, the higher the rotational stellar velocity. Moreover, when the stellar rotation slows down, the lithium depletion increases, leading us to the conclusion above that the stellar rotation is a key element for the lithium-depletion problem.

In summary, we conclude that the low or high Li values are not dependent on the presence of planets, but are associated with the age, mass, and $v\sin{i}$ of the star. Some authors reached similar conclusions \citep[see][and reference herein]{Rom21}, but without considering the presence or absence of planets. To reach these conclusions, we took the influence of tidal forces into account, even in extreme circumstances such as a small orbital separation ($a\leq$ 0.2\,au) and an orbital period shorter than or equal to the stellar rotation period (P$_{\rm orb} \leq$ P$_{\rm rot}$); in this case, for planet to have a high probability of falling onto the star. 
The lithium depletion remains consistent even in these extreme circumstances. 

\section{Lithium depletion and Galactic migration}\label{sec:galactic}

\subsection{Review of high-metal lithium-poor stars}\label{detehmli}

In Sect.~\ref{metallicity}, we mentioned a subset of stars with high metallicity and low lithium abundance. Due to the use of high-resolution spectra, several authors detected this type of stars, e.g. \citet{Del15}, \citet{Gui16}, \citet{Fu2018}, \citet{Ben18}. 
The explanation for the existence of these stars has given rise to the appearance of different models that try to explain how these objects are found in the solar neighbourhood. This means, how these metal-rich objects have lost their lithium well below the value inherited from their natal cloud.  

We considered a homogeneous sample of 1332 field stars (see Sect.\,\ref{sec:sample}), none of which are binaries. 
This implies that our focus is on field stars, and we thereby avoid stellar cluster members and the influence of the internal dynamics of the cluster (see \citealt{Llo22}), which can distort the kinematics of the star. Moreover, other authors such as \citet{Ran20} analysed stellar clusters with high metallicity to state that there is no evidence of an A(Li) decrease at high metallicity. Therefore, cluster stars cannot be mixed with field stars because their primordial values are different \citep{Cha21}. The stars of our sample are located in the thin disc. 

In our sample, 257 are Y stars, that is, $\sim$24$\%$ of the objects in our sample host planets. However, in the case of stars with high metallicity and low lithium abundance, this percentage increases to 42$\%$, with a $\Pi$ ratio of about 1.45 and an average total mass of planets of $\sim$2.36\,M$_{\rm Jup}$. This would make us suspect the appearance of some bias due to, for example, the presence of multiple planets in the system. To verify this, we also calculated the same values, but this time, for stars with high metallicity and high Li abundance (A(Li) $>$ 1.5\,dex). We obtained very similar values: a percentage of 42$\%$ and an average total mass of planets of $\sim$2.56\,M$_{\rm Jup}$. The main difference is in terms of $\Pi$ (about 1.12), of age (5.0$\pm$2.7\,Ga in comparison with 3.00$\pm$1.02\,Ga), and of mass (0.99$\pm$ 0.03\,M$_{\odot}$ in comparison with 1.20$\pm$0.03\,M$_{\odot}$), with an overall mean projected rotational velocity of $\sim$2.4\,km\,s$^{-1}$ (faced to $\sim$4.7\,km\,s$^{-1}$). 
This discussion is summarised in Table\,\ref{table5}, which shows that no biases are associated with the presence and number of planets or their masses. 
Considering the absence of cinematic distortion mentioned earlier, we reaffirm the conclusion described in Sections\,\ref{sec:interaction} and \ref{sec:influence}.

\begin{table}[!h]
\caption{Average values for high-metallicity stars (HM; with [Fe/H] $>$ 0.15\,dex) with low and high lithium abundances.} 
\label{table5}
\centering
\begin{tabular}{lcc}
    \hline\hline 
    & A(Li)<1.0 &  A(Li)>1.5  \\
\hline
NHM [\%] & 42 & 42 \\
$\Pi$ & 1.45 & 1.12 \\
$\Sigma$Planet [Mjup] & 2.36 & 2.56 \\
age [Ga] & 5.0$\pm$2.7 & 3.00$\pm$1.02 \\
$v\sin{i}$ [km\,s$^{-1}$] & 2.4 & 4.7 \\
M [M$_\odot$] & 0.99$\pm$0.03 & 1.20$\pm$0.03 \\
\hline
\end{tabular}
\tablefoot{NHM is the number of HM stars in comparison with our sample of Y stars, given in percent. }
\end{table}

\citet{Lee18} concluded that the evolutionary state of low-Li stars reveals that they are consistent with being evolved Li-dip stars. These authors also suggested based on examining the kinematics and available elemental abundances of the metal-rich stars, that the stars appear to be indistinguishable from lower-metallicity thin-disc stars, except for their high [Fe/H], consistent with an origin in the inner thin disc. 
We also verified whether these low values of the Li abundance could be due to the F dip. We then evaluated it at the stellar effective temperatures and masses: if $T_{\rm eff}>$ 5700\,K and M$\sim$1.4\,M$_{\odot}$, then the action of rotation-induced slow mixing could explain the low A(Li). 
This is because these stars undergo an angular momentum loss and are sufficiently massive for their interior temperatures and densities to be high enough to burn lithium, which in turn could explain the strong lithium depletion \citep{Bau13}.
In our sample, the high-metallicity and low-A(Li) stars have masses between $\sim$0.75 and $\sim$1.68\,M$_{\odot}$ and their $T_{\rm eff}$ lies in between $\sim$4500 and $\sim$6200\,K. However, only two of them (HD\,156098 and HD\,85725) have masses greater than 1.4\,M$_{\odot}$ and $T_{\rm eff}>$ 5700\,K. Therefore, the F-dip lithium scenario is not a plausible explanation for our case. 

In summary, to explain the presence of stars with a high metallicity and low lithium abundance, we would have several possible scenarios or a combination of them. One scenario would be that the depletion is caused by a strong internal process that accelerates the burning of Li related to a decrease in rotation speed, that is, it affects slow rotators \citep{Llo21}, or depletion by the action of stellar accretion discs \citep{Egg12}. Another scenario would be a similar internal process, but acting for a longer time, according to the stellar age \citep{Roc21}. 

Additional explanations could involve depletion caused by other strong internal processes different from those mentioned above, such as a relatively shallow extra mixing \citep[see][]{Mel20} that accelerates the burning of Li related to a decrease in rotation speed, but the depletion of Li should reach a very low value ($\sim$0.5\,dex). Another proposal was described by \citet{Mar23}, who showed that solar twin more metallic stars have deeper convective zones, leading to a more pronounced depletion of lithium compared to stars of lower metallicity. 
We verified this conclusion in our sample and found the same result. However, although our test stars (see Table\,\ref{table6}) have masses similar to the mass of the Sun, they are older stars. The super-solar metallicity causes a deepening of convective zones that would drive a stronger lithium depletion; this strong anti-correlation with metallicity observed in our sample is also within the predictions of non-standard models\citep{Dum21}.  

Another scenario that contemplates the old age of our candidates would be the processes described above, either isolated or jointly, but acting for a longer time, which might explain the strong Li depletion. \citet{Gui16,Gui19} proposed that metal-rich dwarf stars are indeed old stars, but come from the inner regions of the Galaxy. These stars would have depleted their lithium in their journey from their place of birth, which had a high metallicity, to their current position.
Both the phenomenology that we proposed in the previous paragraph and in this last scenario led us to consider their action during the trajectory time from their places of origin to the surroundings of the Sun; their current position. This is the topic is addressed in the next subsection.

\begin{figure}[!ht]
    \centering
        \includegraphics[width=.98 \columnwidth]{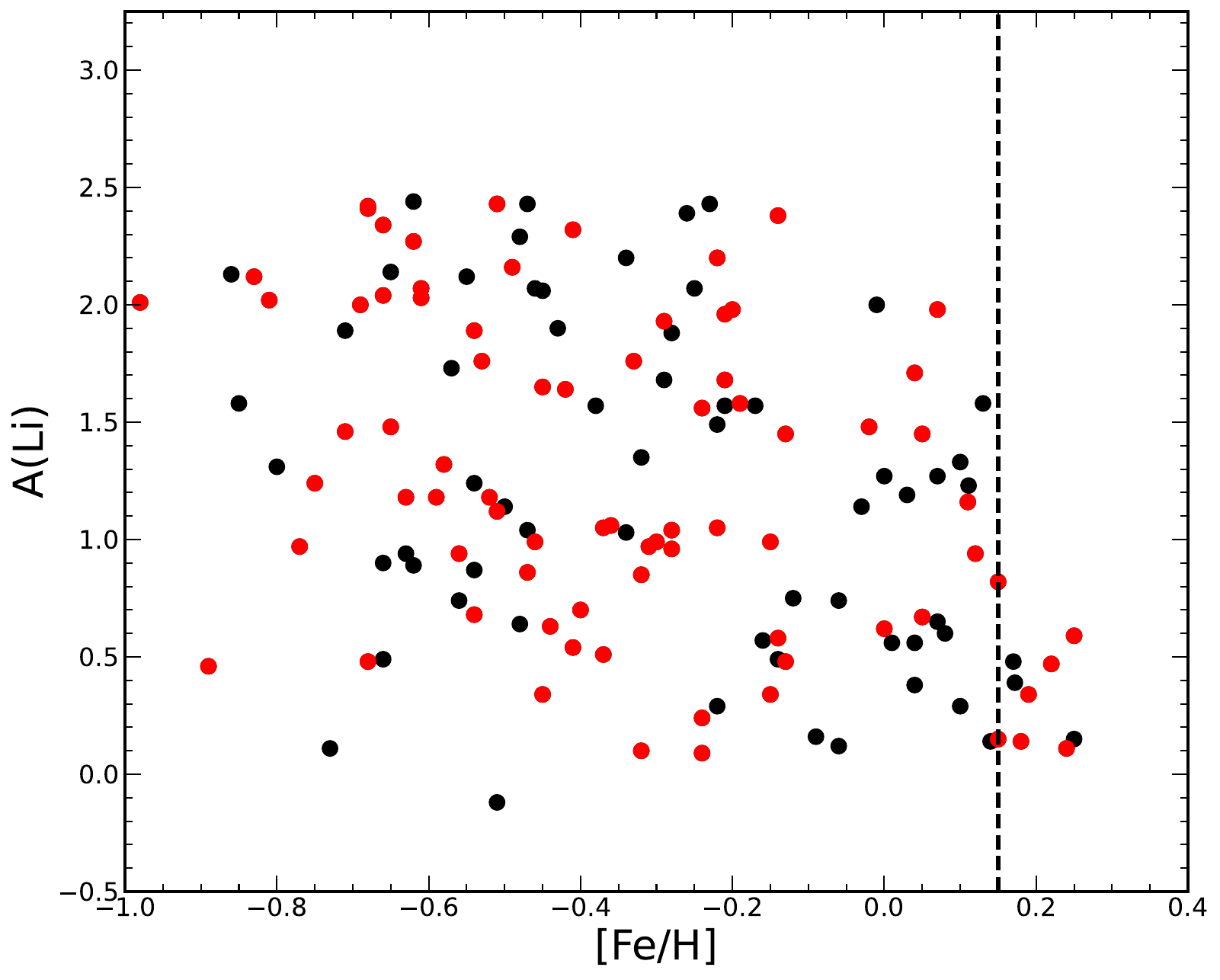}
    \caption{Abundance of lithium as a function of metallicity for stars with ages $\geq 8 \pm 1.3$\,Ga. We highlight stars with negative values of $U$ in red. High-metallicity stars with very low A(Li) are shown on the right side of the dashed line.}
     \label{figure6}
\end{figure}

\begin{figure}[!ht]
    \centering
    \includegraphics[width=.99 \columnwidth]{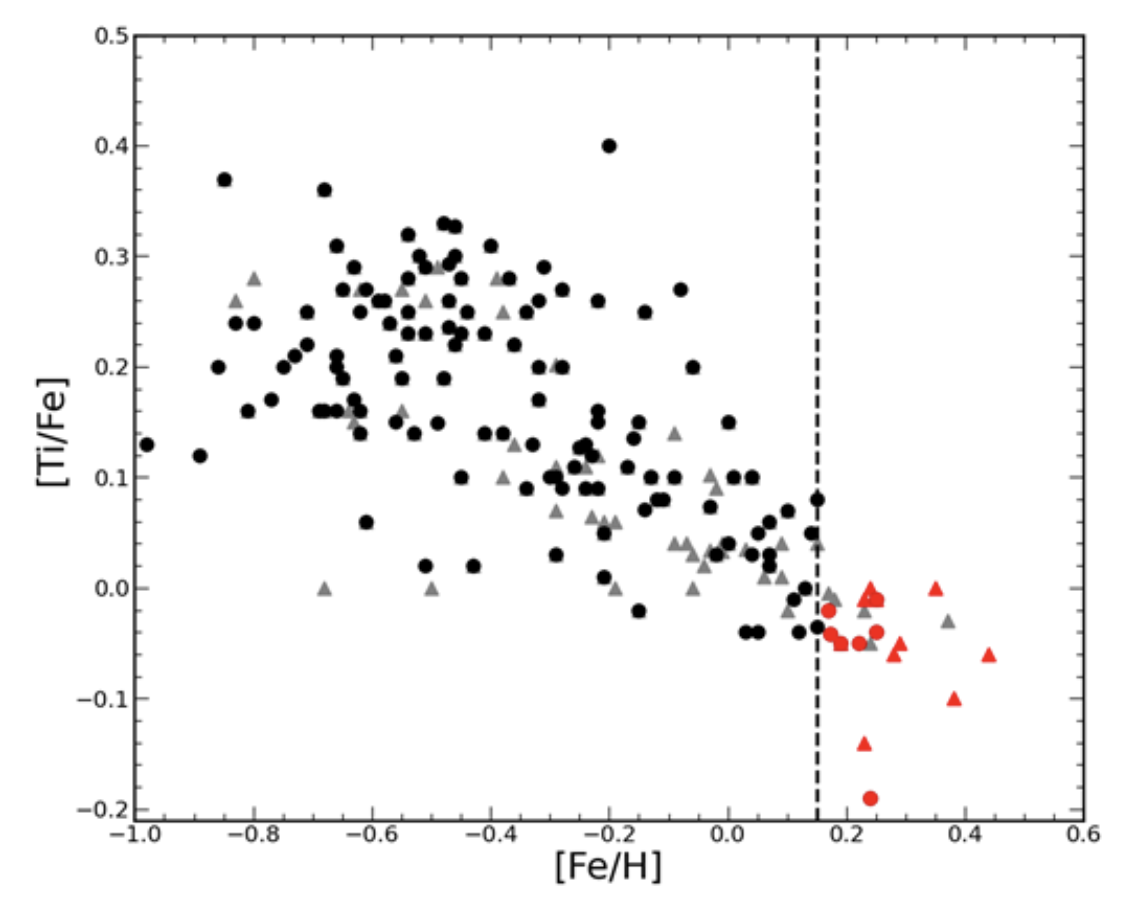}
    \caption{Relation of [Ti/Fe] with respect to [Fe/H] for all stars with an age greater than 8\, Ga. The black dots represent stellar ages computed using the PARAM code, and grey triangles represent stars with an age quoted from the literature. Stars with super-solar metallicity ([Fe/H]> 0.15\,dex) and a low lithium abundance (A(Li) <1.0\,dex) are located on the right side of the dashed line and are identified by full symbols in red. A detailed explanation can be found in the text.}
     \label{figure7}
\end{figure}


\subsection{Evaluation of the migration model}\label{modonmi}

Some scenarios of internal processes were described in the previous subsection to explain the strong Li depletion in stars in the solar neighbourhood with a high metallicity. 
Whether these processes act in isolation or in tandem, an elapsed time is necessary to overcome the two main constraints: high metallicity, and the stellar position in the solar neighbourhood. In this section, we analyse whether this required time aligns with the time the star spends on its journey from its birthplace to its current location in the solar vicinity. 

An interesting model to explain the origin of these high-metallicity and lithium-poor stars is the scenario proposed by \citet{Gui16,Gui19}, which suggests that these stars originated in the central metal-rich regions of the Galaxy, migrated to the solar vicinity, and depleted their lithium content during the trip. 
\citet{Cha21} found based on cool dwarf stars that the A(Li) decreases at super-solar metallicities, which obscures the actual Li evolution picture. The authors concluded that field stars and open cluster surveys with an opposite Li behaviour in the high-metallicity regime did not sample the same types of stars. In consequence, they suggested that the original Li abundance in metal-rich Galactic environments could be even higher than the mean value derived in their work. However, their sample contained fast rotators, and the Li depletion accordingly increases when the projected rotational velocity reaches its critical value, which in turn depends on the age, as shown in LA21.

Recently, \citet{Dan22,Dan23} studied this problem and supported the hypothesis of \citet{Gui19}. Based on a sample of 1460 field stars and their corresponding chemical composition, the authors concluded that stars with a high metallicity in their place of birth are incompatible with any site in the solar neighbourhood. Thus, these stars are expected to have travelled from other regions in the inner part of the Galaxy to the solar vicinity. \citet{Del23} also recently considered a situation that is compatible with the mentioned scenario. They showed based on the relation between A(Li) and [Fe/H] two branches, one of lithium-rich (with A(Li) > 1.5\,dex) stars and the other of lithium-poor (A(Li) < 1.0\,dex) stars \citep[][fig.\,5]{Del23}. The second branch includes a few high-metallicity stars (with [Fe/ H] $\ge$ 0.15\,dex) with very low A(Li) values of around 0.5\,dex. 
Among them, those stars that have negative Galactic velocities ($U$), meaning that they came from the central Galactic regions, are good candidates for an ulterior more complete kinematic or dynamical study. Due to their poor statistics (46 stars), the authors were not able to make a strong statement about the validity of the scenario presented by \citet{Gui19}.

To identify the best objects for our purpose, those with high metallicity and a low lithium abundance, we split our sample of stars into two: stars younger and older than 8\,Ga. The latter is displayed in Fig.~\ref{figure6}. Stars with a negative Galactic velocity are shown as red dots. The super-metal stars with very-low A(Li) are located to the right of the dashed line. 

The choice of the adopted age may be relevant to the conclusions. We therefore carried out our investigation with age values from the PARAM code, but also used the age extracted from the literature. Both values are described in LA21 and listed in their Table 1. The mean error in the age is $\sim$1.5\,Ga, but in the case of age from the literature, we eliminated stars with errors larger than 2\,Ga to avoid any distortion caused by a larger error, which introduces a high degree of dispersion in our results (most of them came from K-type stars).

Our stars belong to the thin Galactic disc. To ensure that we used the element Ti, which is a known proxy for alpha elements \citep{Del23}. These values were calculated using the relative iron abundances ([Fe/H]) and [Ti/H] values taken from the literature. For stars without an available [Ti/H] value, we quote the value of [$\alpha$/Fe] from the literature. The adopted values as well as the literature references for [Ti/H] and [$\alpha$/Fe], when applied, are listed in Table\,\ref{tableA2}. The relation of [Ti/Fe] versus [Fe/H] for stars older than 8\,Ga, either computed with the PARAM code (represented by dots) or taken from the literature (represented by triangles), is depicted in Fig.\,\ref{figure7}. This figure shows that the majority of objects are distributed in the low-metallicity region, with [Fe/H] < $-$0.1\,dex. Their corresponding [Ti/Fe] values are higher than $\sim$0.04\,dex, chemically indicating that they belong to the thin disc (see e.g. \citealt{Del23}). This means that our target stars belong to the thin disc.

\begin{table}[!ht]
\caption{Stars illustrated in Fig.~\ref{figure7} as having a high metallicity and low lithium abundance.} 
\label{table6}
\centering
\begin{tabular}{lcccc}
\hline
\hline   
   Identifier & age$_{\rm PARAM}$  & age$_{\rm Lit}$ & $U$ & Candidates \\
    & (Ga)  & (Ga) & km\,s$^{-1}$ &  \\ 
\hline

HD 166745	&	8.2$\pm$1.6	&	\ldots	& -7.61  &   $\bullet$	\\
HD 190647	&	8.9$\pm$1.0  &   8.0$\pm$1.2 & -32.37	&	$\bullet$	\\
HD 22177 	&	8.2$\pm$1.6  & 8.1$\pm$1.4   & -8.89 &	$\bullet$	\\
HD 30306	    &	8.4$\pm$1.7   & 8.5$\pm$2.7   &	5.00 &	\ldots	\\
HD 45350 	&	8.1$\pm$1.5  & 9.2$\pm$1.6   &	16.65   &	\ldots	\\
HD 64640	    &	9.5$\pm$1.7   &   \ldots  &   -23.83  &	$\bullet$	\\
HD 73526	    &	8.1$\pm$1.8   & \ldots    & -78.91  &	$\bullet$	\\
Kepler 46	&	10.5$\pm$1.2    & 11.8  & 258.74   &	\ldots	\\
HD 102117	&	\ldots	&	8.1$\pm$1.4 & 13.35	&	\ldots	\\
HD 108874	&	\ldots	&	12.3$\pm$1.0 &	44.91	&	\ldots	\\
HD 115585	&	\ldots	&	8.5$\pm$1.2 &	-50.14	&	$\bullet$	\\
HD 117207	&	\ldots	&	8.4$\pm$1.6 &	-33.40	&	$\bullet$	\\
HD 134606	&	\ldots	&	12.2$\pm$1.1 &	-20.98	&	$\bullet$	\\
HD 145675	&	\ldots	&	12.3$\pm$0.9 &	23.35	&	\ldots	\\
HD 46375	    &	\ldots	&	10.0$\pm$1.4 &	9.22	&	\ldots	\\
HD 76909	    &	\ldots	&	8.4$\pm$1.7 &	24.72	&	\ldots	\\

 \hline 
  
\end{tabular}                               
\tablefoot{Both age estimates, either from the PARAM code or from the literature, are presented. The last column lists objects that are good candidates for validating the scenario proposed by \citet{Gui19}.}
                             
\end{table}

The two samples of stars with different age estimates, those from the PARAM code and from the literature, are depicted in Fig.\,\ref{figure7}, where they are illustrated as red dots and red triangles, respectively. They are also listed in Table\,\ref{table6}. 
Figure\,\ref{figure7} also shows on the right side of the dashed line a few stars with a high Li abundance (grey triangles), mixed with the high-metallicity and low-Li stars (red symbols). Their presence could be explained by the difference between the two sources of age estimates. These objects are HD\,96167 (with A(Li)$\sim$1.65\,dex, age$_{\rm PARAM}$$\sim$4.1\,Ga); HD\,127334 (with A(Li)$\sim$1.61\,dex, age$_{\rm PARAM}$$\sim$7.1\,Ga); HD\,43691 (with A(Li)$\sim$2.21\,dex, age$_{\rm PARAM}$$\sim$2.9\,Ga); HD\,179949 (with A(Li)$\sim$2.50\,dex, age$_{\rm PARAM}$$\sim$1.1\,Ga); and HD\,8648 (with A(Li)$\sim$1.49\,dex, age$_{\rm PARAM}$$\sim$6.9\,Ga). These stars were not considered.

The stars listed in the last column of Table\,\ref{table6} would be the best test stars, regardless of the adopted age, to validate the migration scenario proposed by \citet{Gui19} because they are thin-disc stars older than 8\,Ga, which would have spent enough time to deplete their lithium along their trip from their birthplace until their current positions in the solar vicinity. They are suggested to come from the inner part of the Galaxy. They include stars with and without planets.

We studied the migration history of the stars marked in Table~\ref{table6} to discuss whether they might have depleted their lithium over their Galactic trajectory. 
These stars have an average age obtained with the PARAM code of 8.7$\pm$1.5\,Ga and an average age from the literature of 10$\pm$1.3\,Ga. The stars HD\,166745, HD\,190647 HD\,22177, HD\,64640, and HD\,73526 have a negative Galactic velocity $U$, as shown in Table~\ref{table6}. Of those stars that present positive velocities, Kepler\,46 has the higher value ($U=$ 258.74\,km\,s$^{-1}$). It is also one of the oldest stars. The stars with ages gathered from the literature and with negative $U$ velocities are HD\,115585, HD\,117207, HD\,134606, HD\,190360, and HD\,22177. Of the stars with ages from the literature and positive velocities, HD\,108874 is the oldest star.

All stars displayed in Table \ref{table6} either with $U$<0 or $U$>0\,km\,s$^{-1}$ or with age PARAM or age literature are very slow rotators, with projected rotational velocities slower than 2\,km\,s$^{-1}$. From this critical projected rotational velocity, the Li abundance values decrease rapidly (LA21), which does not contribute to a possible shortening of the length of the convection layer and, therefore, to a greater speed in the burning of lithium. We note that for the ages estimated using the PARAM code, the maximum value for the Li abundance is A(Li) $\leq$ 0.60\,dex, but for ages obtained from the literature, the corresponding maximum value is A(Li) $\leq$ 0.91\,dex.  
They also present a low magnetic activity, which is $\log{R'_{\rm HK}}$ $\sim$$-$5.0 on average, except for HD\,30306, which is very active ($\log{R'_{\rm HK}}$ $\sim$$-$4.5). The average for the metallicity is approximately 0.205\,dex and $\sim$0.305\,dex in the case of samples with ages obtained with the PARAM code and from the literature, respectively. Their masses are about 1$\pm$0.02M$_{\odot}$, which does not contribute to a possible shortening of the convection zone.

We remark that the fact that all stars are very slow rotators represents a very favourable condition for any future analysis of the lithium depletion in stars migrating from the inner regions of the Galaxy. Firstly, this condition avoids any complications related to faster stellar rotational velocities, with $v\sin{i}$ values higher than the critical velocity of 5\,km\,s$^{-1}$, which in turn retards the lithium depletion. This critical velocity has been discussed in LA21. Secondly, as noted by \citet{Gui16,Gui19}, if these stars come from inner metal-rich regions, studies of the stellar lithium depletion would require specific theoretical models of this internal depletion under high-metallicity conditions.

This caused us to study the kinematic behaviour of these stars. To analyse the kinematics of the stars presented in Table\,\ref{table6}, we employed {\it galpy}, a Python package for galactic dynamics calculations fully described in \citet{Bov15}. When a Galactic potential is selected, this tool can calculate the trajectory of a star in the Milky Way, taking its position, proper motion, radial velocity, and distance, as well as the distance to the Galactic centre from the Sun (R$_0$) and the circular velocity in that point (v$_0$) as inputs. The proper motions, radial velocities, and distances for our sample of high-metallicity Li-poor stars were taken from {\it Gaia} DR3 \citep{gaiaDR3}. Despite the plethora of data released in Gaia DR3, HD\,73526 and Kepler\,46 do not have radial velocities available in that source. We therefore turned to \citet{GC18} and \citet{Bre18}, respectively, to obtain this information. We employed three different potentials for our analysis: the potential described in \citet{Bov15} (hereafter MW2014); the galactic potential described in \citet{Cau20} (hereafter Cautun2020); and the MW2014 potential modified by adding the bar described in \citet{Deh00} (hereafter MW2014+DW). Because the results were similar in the three cases, we only discuss the results obtained with the MW2014+DW potential and the R$_0$ and v$_0$ taken from \citet{Pod23}.

Table\,\ref{galpy_results} shows the metallicity, age, Li abundance, and the radial component of the galactocentric velocity ($U$) for our sample of 16 high-metallicity Li-poor stars, as well as the minimum and maximum distances to the Galactic centre and the orbit eccentricities derived from our analysis with {\it galpy}. Even though this list of stars is equally divided in velocities $U$<0 and $U$>0\,km\,s$^{-1}$, the results appear to be independent of the Li abundances. This means that completely different Galactic dynamic histories provide the required low Li abundances.

\begin{figure}
   \centering
   \includegraphics[width=.99 \columnwidth]{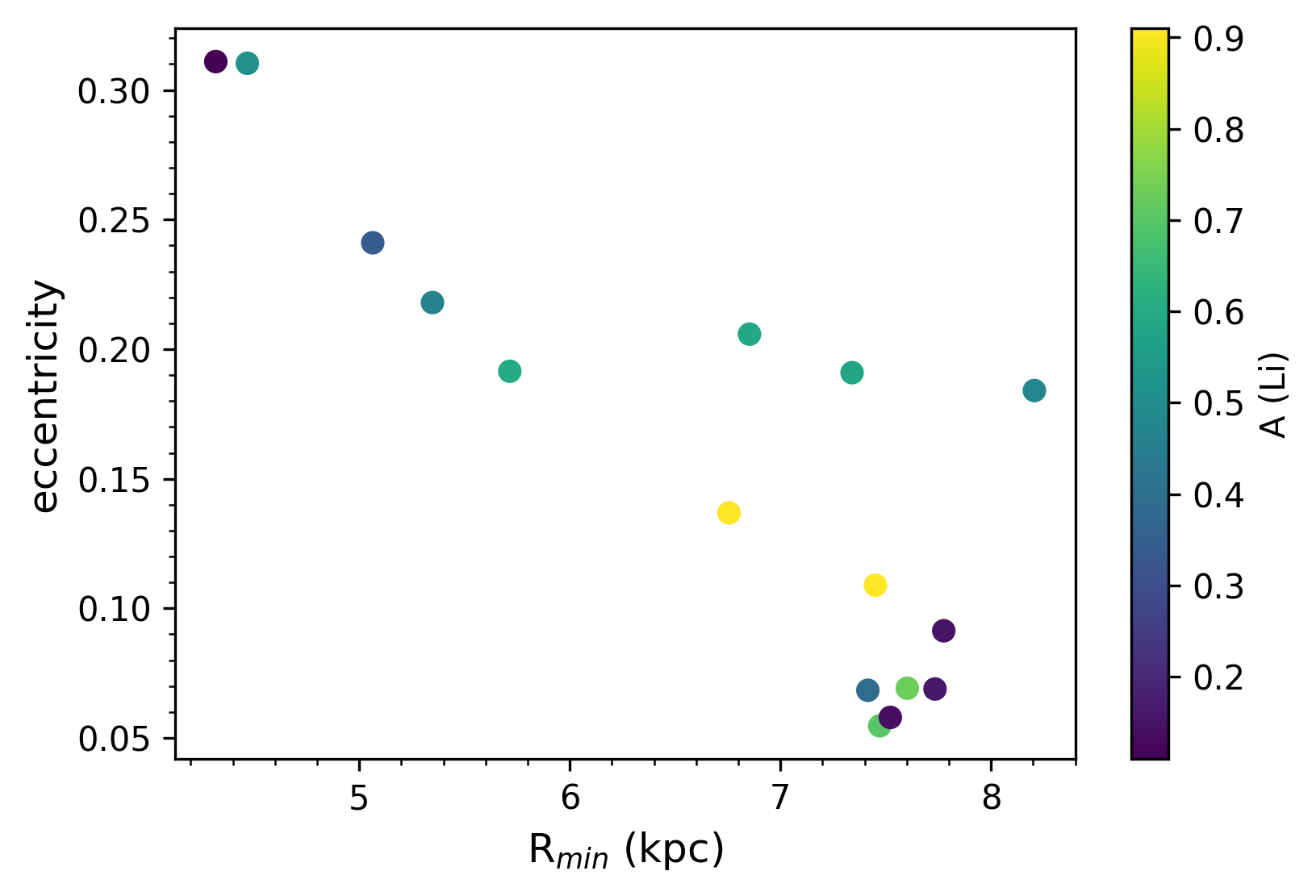}
   \caption{Eccentricity and minimum distance to the Galactic centre obtained with {\it galpy} for the 16 high-metallicity Li-poor objects listed in Table \ref{table6}. The symbol colours indicate the lithium abundances listed in Table \ref{galpy_results}.} 
        \label{Rmin_ALi}%
\end{figure}

\begin{figure}
    \centering
	\includegraphics[width=.99 \columnwidth]{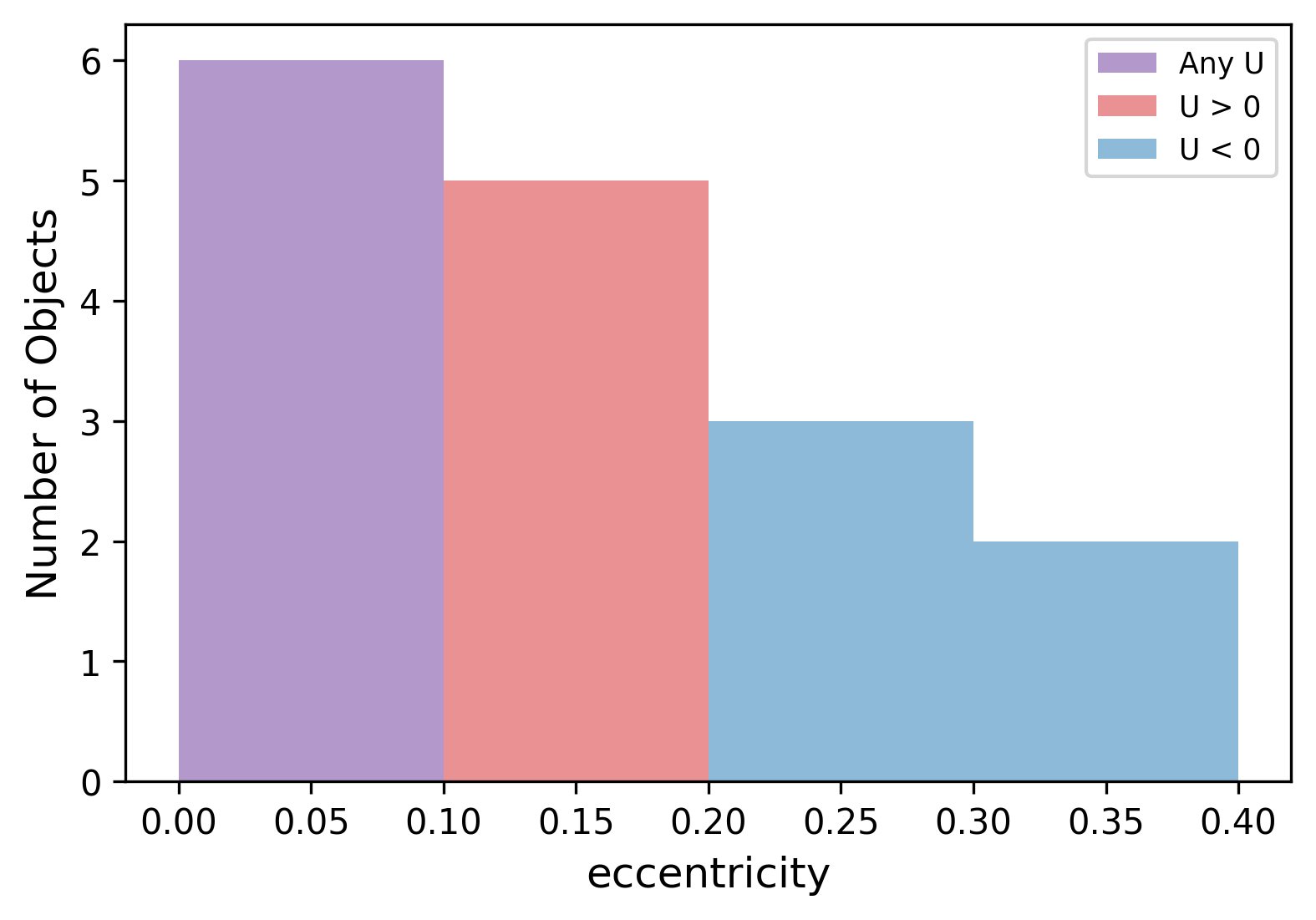}
	\caption{Histogram showing the orbit eccentricities obtained with {\it galpy} for the 16 high-metallicity Li-poor objects listed in Table\,\ref{table6}. The positive values of the radial component of the Galactocentric velocity are clearly correlated with eccentricities below 0.20. The purple bar between e$=$0.00 and e$=$0.10 represents three objects with $U$<0 together with three objects with $U$>0\,km\,s$^{-1}$.}
		\label{e_hist}
\end{figure}

The minimum distances to the Galactic centre and the Li abundances are presented in Fig.\,\ref{Rmin_ALi}, which shows an important result of our dynamical calculations regarding the distributions of orbital eccentricities. 
Higher eccentricities appear to be associated with negative $U$ values (Fig.\,\ref{e_hist}). In particular, stars with eccentricities greater than 0.2 are closer to the Galactic centre. 
Although HD\,117207, HD\,134606, and HD\,64640, which are stars with negative $U$ values, have eccentricities below 0.10, and considering their low Li abundances, their obtained R$_{min}$ is over 7.0\,kpc (see Fig.\,\ref{Rmin_ALi}), the members of our sample appear to deviate from the chemical models presented in \citet{Gui19}. 
Fig.\,\ref{Rmin_ALi} shows that most of the sample exhibit R$_{min}$ > 6\,kpc, with only five stars getting closer to the Milky Way centre (R$_{min}$ < 6\,kpc).

\begin{table*}
\caption{High-metallicity Li-poor stars in our sample. } 
\label{galpy_results}      
\centering          
\begin{tabular}{c c c c c c c c}     
\hline\hline        
Identifier & Age\tablefootmark{a} (Ga) & [Fe/H] & A(Li) & U (km\,s$^{-1}$) & R$_{min}$ (kpc) & R$_{max}$ (kpc) & e \\ 
\hline                     
   HD115585 & 8.5$\pm$1.2 & 0.35 & 0.51 & -50.14 & 4.47 & 8.49 & 0.31 \\
   HD117207 & 8.4$\pm$1.6 & 0.23 & 0.16 & -33.4 & 7.73 & 8.88 & 0.07 \\
   HD134606 & 12.2$\pm$1.1 & 0.28 & 0.7 & -20.98 & 7.47 & 8.34 & 0.06 \\
   HD166745 & 8.2$\pm$1.6 & 0.24 & < 0.11 & -7.61 & 4.32 & 8.22 & 0.31 \\
   HD190647 & 8.9$\pm$1.0 & 0.22 & < 0.47 & -32.37 & 5.35 & 8.33 & 0.22 \\
   HD22177 & 8.2$\pm$1.6 & 0.19 & < 0.34 & -8.89 & 5.07 & 8.28 & 0.24 \\
   HD64640 & 9.5$\pm$1.7 & 0.18 & < 0.14 & -23.83 & 7.52 & 8.45 & 0.06 \\
   HD73526 & 8.1$\pm$1.8 & 0.25 & 0.59 & -78.91 & 6.85 & 10.41 & 0.21 \\
   HD102117 & 8.1$\pm$1.4 & 0.29 & 0.6 & 13.35 & 5.72 & 8.42 & 0.19 \\
   HD108874 & 12.3$\pm$1.0 & 0.23 & 0.58 & 44.91 & 7.34 & 10.81 & 0.19 \\
   HD145675 & 12.3$\pm$0.9 & 0.44 & 0.91 & 23.35 & 7.45 & 9.27 & 0.11 \\
   HD30306 & 8.4$\pm$1.7 & 0.17 & < 0.48 & 5.0 & 8.21 & 11.91 & 0.18 \\
   HD45350 & 8.1$\pm$1.5 & 0.25 & < 0.15 & 16.65 & 7.78 & 9.34 & 0.09 \\
   HD46375 & 10.0$\pm$1.4 & 0.24 & 0.73 & 9.22 & 7.60 & 8.73 & 0.07 \\
   HD76909 & 8.4$\pm$1.7 & 0.38 & < 0.91 & 24.72 & 6.76 & 8.90 & 0.14 \\
   Kepler-46 & 10.5$\pm$1.2 & 0.172 & 0.39 & 258.74 & 7.42 & 8.50 & 0.07 \\
   
\hline                  
\end{tabular}
\tablefoot{The age, metallicity, Li abundance, and the $U$ component of the Galactocentric velocity are taken from LA21. The minimum (R$_{min}$) and maximum (R$_{max}$) distance to the Galactic centre and the orbit eccentricity ($e$), obtained with the MW2014+DW potential, and the R$_0$ and v$_0$ taken from \citet{Pod23}, are also listed for each object.\\
\tablefoottext{a}{Age sources are explained in the text. }
}
\end{table*}

%

To summarise, based on our kinematics results, we deduce that neither the minimum values of the galactocentric radius nor the eccentricities allow us to conclude that low Li abundances associated with high metallicities come from an enrichment in metals due to the journey of the star through the inner regions of the Galaxy. It might be concluded, however, that the scenario proposed by \citet{Gui16,Gui19} is insufficient to explain the existence of these stars in the solar neighbourhood.

For the lithium abundance to decrease to a level of A(Li)$\sim$0.6\,dex, stars must have undergone a significant intrinsic depletion. \citet{Mar23} proposed that super-solar metallicity stars show a deepening of their convective zones, resulting in a strong lithium depletion, which agrees with the predictions of non-standard models \cite{Dum21}. We suggest that these strong internal processes would need to either accelerate the convection process, despite being slow rotators, to bring lithium to the burning zone or remain active for an extend period. However, this period must be shorter than the time elapsed during their journey from the birthplace to the present position.

From our previous studies, CH19, LA21, RF21, and based on the results by \citet{Del23}, we might also propose that these stars were formed from [Fe/H]-rich gas. They become slow rotators due to large stellar accretion discs and, in consequence, they accelerate the Li depletion process. Even in this case, these stars would not have had enough time to reduce their lithium content to the current values because the younger the stars, the weaker the lithium depletion. 
In the future, upcoming data from large surveys such as GALAH \citep{Bud21} combined with better determinations of A(Li) in Li-poor dwarf stars will result in unbiased large samples of main-sequence stars. These new catalogues will allow us to better understand the many channels of Li production or destruction that occur inside dwarf stars.

\section{Conclusions} \label{conclusions}

We have addressed the enduring debate of whether solar mass planet-hosting stars have different lithium abundances than stars without detected planets. 
We analysed a sample of 1332 stars, including 257 planet-hosting stars, and found that the presence of planets has no clear influence on the measured lithium abundances.
We examined whether a planet merger enriches the host stars with lithium.
Additionally, we investigated how various stellar properties such as metallicity, activity, and rotational velocities affect the lithium abundances with the aim to discern potential differences between stars with and without planetary companions.

We analysed the influence of planet mergers using two methods: 1) We calculated the orbital decay for a subgroup of 34 stars with the highest probability of experiencing a final spiralling of a planet onto its host star, and 2) we used models by \citet{Sev22} to examine lithium abundance enrichment signatures after planetary engulfment. Masses between 0.7 and 1.0\,M$_{\rm \odot}$ favour the longest new lithium lifetimes, of about 1\,Ga or longer. 
However, the distribution of stellar lithium abundances in host stars, whose stellar masses are between 0.7 and 1.5\,M$_{\rm \odot}$ and that have low OD values do not indicate any lithium enrichment. We therefore conclude that for the merging of planets with masses of 1, 10, and 100\,M$_{\rm \oplus}$, which are those considered in the models of \citet{Sev22}, no lithium contamination appears to be present in the studied solar mass stars. 
Even when considering extreme orbital conditions of planets, such as those in close-in orbits and with an orbital period shorter than the stellar rotation period (in which case the mutual influence between stars and planets is strongest), we also found that there is no difference in the behaviour of lithium in stars with and without planets. 

FGK-dwarf stars with planets on close orbits, particularly those with multi-planet or single systems with $\Sigma M_p \geq$ 1\ M$_{\rm Jup}$, tend to have higher metallicity than those without detected planets. 
However, the relation between metallicity and the proportion of close-orbiting planets varies, with a higher proportion observed for lower metallicity systems with $\Sigma M_p <$ 1\,M$_{\rm Jup}$.

We found that stars with A(Li) $\geq$ 1.5\,dex show a higher probability to have intense chromospheric activity than those with A(Li) $<$ 1.5\,dex. 
This also correlates with the stellar rotational velocity. Fast-rotating stars have a higher chromospheric activity and higher A(Li) \citep{Roc21}. 
We did not observe a relation between the presence of planets and the characteristics we analysed. However, our analysis falls in line with \citet{Llo21}, providing further evidence of the tight lithium-rotation connection.

The migration of stars from central regions of the Galaxy up to the solar vicinity is a complex subject. It is incompletely resolved how lithium in dwarf stars, originally from high-metallicity central zones, has been depleted and transformed into the observed low lithium abundances in nearby stars. 
There are several reasons: The travel times of individual stars, which are essential for understanding the depletion mechanisms during their journey, are generally unknown. 
Moreover, theoretical models of lithium depletion for super-metal stars are not yet available, which complicates the problem. Some authors such as \citet{Gui19} and \citet{Dan22} have tackled the migration processes using Galactic chemical models in which stars are considered as flows of stars crossing Galactic regions between the central regions and the solar neighbourhood. 
The primary parameters considered by these authors to support the argument for a central Galactic origin of these stars were age and metallicity.

In contrast to previous studies that focused on stellar currents, we examined individual stars with known parameters that may have migrated from the central Galactic regions. 
While \citealt{Del23} made an initial attempt with a limited number of stars that had both planets and debris discs, we expanded on this approach by examining 1075 stars without detected planets and 257 planet-host stars, all with measured lithium abundances.
Our results indicate that the scenario proposed by \citet{Gui19} cannot fully account for stars with high metallicity and a low lithium abundance. To reach low A(Li) values of around 0.5\,dex, stars must have a significant intrinsic depletion due to their high metallicities. 
Furthermore, the trajectories built for these objects from their current positions, proper motions and radial velocities diverge from the chemical models proposed in \citet{Gui19}. The results for either for $U$<0 or $U$>0\,km\,s$^{-1}$ appear to be independent of the Li abundances. Moreover, stars with high eccentricities (e>0.2) reach the shortest distances to the Galactic centre. This means that different Galactic dynamic stories would have caused the required low Li abundances. 
It seems more appropriate that the depletion is caused by processes that accelerate the burning of Li either in very slow rotators or related to a decrease in rotation speed, such as the action of stellar accretion discs. 
In addition, the explanation proposed by \citet{Lee18}, who suggested that low-Li stars could be evolved F-dip stars, is also insufficient to explain the observed low A(Li). However, in our subsample, the stellar masses are below 1.1 M$_\odot$, which is lower than the expected mass for F lithium-dip stars.

Moreover, even in the case of strong accelerating internal process, this process has to be shorter than the travel time from the birthplaces to the present position. Based on our previous works, we can also propose that due to galactic-scale events, some stars formed from [Fe/H]-rich region, becoming slow rotators. These stars would not have had enough time to diminish their lithium content to the current values.

Lithium is a very sensitive element and can be affected by various factors.
We found no clear influence of the presence of planets on the lithium abundances in solar mass stars, even in extreme orbital conditions.
However, a higher metallicity and fast rotation were observed in stars with a high lithium abundance.
In addition, the low lithium abundances of the high-metallicity Li-poor stars in our sample cannot be explained by solar Li-depletion alone.

To summarise, the exact mechanisms behind lithium depletion remain uncertain and require further study.
This work aims to serve as a gateway for future proposals to increase the database of lithium-depleted stars and explore their origin using ground- and space-based facilities. 
A larger sample of detected planets on close-in orbits will help us to better understand the effects of orbital decay, stellar activity, and lithium abundances. 
Furthermore, a dedicated study of a set of high-metal low-Li stars is needed to better understand the migration-depletion scenario.

\section*{Acknowledgements}

We would like to express our gratitude to an unknown referee for his comments and suggestions. 
The authors thank  Dr S. Roca-F\`abrega his constructive comments and helpful insights.
FLA and PC acknowledge the support from the Faculty of the European Space Astronomy Centre (ESAC), under funding reference number ESA-SCI-SC-LE-059. 
FLA also would like to thank the technical support provided by A. Parras (CAB), Dr J. A. Prieto (UCLM), and MSct.P. Cabo for her support in the use of the {\tt SVODiscTool}.
CC acknowledges financial support from the ERDF through the project AYA2016-79425-C3-1/2/3-P. 
PC acknowledges financial support from the Government of Comunidad Autónoma de Madrid (Spain) via the postdoctoral grant `Atracción de Talento Investigador' 2019-T2/TIC-14760. PC also thanks Dr M. G\'alvez-Ortiz and Dr E. Solano for the fruitful discussion and support. 
DCM acknowledges financial support from the Spanish State Research Agency (AEI) Projects MCIN/AEI/10.13039/501100011033 grant PID2019-107061GB-C61 and No. MDM-2017-0737 Unidad de Excelencia “María de Maeztu”- Centro de Astrobiología (INTA-CSIC). DCM also thanks J.L. Gragera-Mas (CAB) for his comments and Dr D. Barrado.
E. J. Alfaro acknowledges financial support from the State Agency for Research of the Spanish MCIU through the "Center of Excellence Severo Ochoa" award to the Instituto de Astrofísica de Andalucía (SEV-2017-0709),  and from MCIU PID2022-136640NB-C21 grant.
This research has made use of the Spanish Virtual Observatory (\url{https://svo.cab.inta-csic.es}) project funded by MCIN/AEI/10.13039/501100011033/ through grant PID2020-112949GB-I00.

\bibliographystyle{aa} 
\bibliography{main} 

\begin{thebibliography}{86}
\expandafter\ifx\csname natexlab\endcsname\relax\def\natexlab#1{#1}\fi

\bibitem[{{Aguilera-G{\'o}mez} {et~al.}(2018){Aguilera-G{\'o}mez}, {Ram{\'\i}rez}, \& {Chanam{\'e}}}]{Agu18}
{Aguilera-G{\'o}mez}, C., {Ram{\'\i}rez}, I., \& {Chanam{\'e}}, J. 2018, \aap, 614, A55

\bibitem[{{Barker}(2020)}]{Bar20}
{Barker}, A.~J. 2020, \mnras, 498, 2270

\bibitem[{{Barker} \& {Ogilvie}(2009)}]{Bar09}
{Barker}, A.~J. \& {Ogilvie}, G.~I. 2009, \mnras, 395, 2268

\bibitem[{{Barker} \& {Ogilvie}(2011)}]{Bar11}
{Barker}, A.~J. \& {Ogilvie}, G.~I. 2011, \mnras, 417, 745

\bibitem[{{Baugh} {et~al.}(2013){Baugh}, {King}, {Deliyannis}, \& {Boesgaard}}]{Bau13}
{Baugh}, P., {King}, J.~R., {Deliyannis}, C.~P., \& {Boesgaard}, A.~M. 2013, \pasp, 125, 753

\bibitem[{{Baumann} {et~al.}(2010){Baumann}, {Ram{\'\i}rez}, {Mel{\'e}ndez}, {Asplund}, \& {Lind}}]{Bau10}
{Baumann}, P., {Ram{\'\i}rez}, I., {Mel{\'e}ndez}, J., {Asplund}, M., \& {Lind}, K. 2010, \aap, 519, A87

\bibitem[{{Behmard} {et~al.}(2022){Behmard}, {Dai}, {Brewer}, {Berger}, \& {Howard}}]{Beh22}
{Behmard}, A., {Dai}, F., {Brewer}, J.~M., {Berger}, T.~A., \& {Howard}, A.~W. 2022, arXiv e-prints, arXiv:2210.12121

\bibitem[{{Behmard} {et~al.}(2023){Behmard}, {Sevilla}, \& {Fuller}}]{Beh23}
{Behmard}, A., {Sevilla}, J., \& {Fuller}, J. 2023, \mnras, 518, 5465

\bibitem[{{Bensby} {et~al.}(2014){Bensby}, {Feltzing}, \& {Oey}}]{Ben14}
{Bensby}, T., {Feltzing}, S., \& {Oey}, M.~S. 2014, \aap, 562, A71

\bibitem[{{Bensby} \& {Lind}(2018)}]{Ben18}
{Bensby}, T. \& {Lind}, K. 2018, \aap, 615, A151

\bibitem[{{Bovy}(2015)}]{Bov15}
{Bovy}, J. 2015, \apjs, 216, 29

\bibitem[{{Brewer} \& {Fischer}(2018)}]{Bre18}
{Brewer}, J.~M. \& {Fischer}, D.~A. 2018, \apjs, 237, 38

\bibitem[{{Buder} {et~al.}(2021){Buder}, {Sharma}, {Kos}, {Amarsi}, {Nordlander}, {Lind}, {Martell}, {Asplund}, {Bland-Hawthorn}, {Casey}, {de Silva}, {D'Orazi}, {Freeman}, {Hayden}, {Lewis}, {Lin}, {Schlesinger}, {Simpson}, {Stello}, {Zucker}, {Zwitter}, {Beeson}, {Buck}, {Casagrande}, {Clark}, {{\v{C}}otar}, {da Costa}, {de Grijs}, {Feuillet}, {Horner}, {Kafle}, {Khanna}, {Kobayashi}, {Liu}, {Montet}, {Nandakumar}, {Nataf}, {Ness}, {Spina}, {Tepper-Garc{\'\i}a}, {Ting}, {Traven}, {Vogrin{\v{c}}i{\v{c}}}, {Wittenmyer}, {Wyse}, {{\v{Z}}erjal}, \& {GALAH Collaboration}}]{Bud21}
{Buder}, S., {Sharma}, S., {Kos}, J., {et~al.} 2021, \mnras, 506, 150

\bibitem[{{Castro} {et~al.}(2009){Castro}, {Vauclair}, {Richard}, \& {Santos}}]{Cas09}
{Castro}, M., {Vauclair}, S., {Richard}, O., \& {Santos}, N.~C. 2009, \aap, 494, 663

\bibitem[{{Cautun} {et~al.}(2020){Cautun}, {Ben{\'\i}tez-Llambay}, {Deason}, {Frenk}, {Fattahi}, {G{\'o}mez}, {Grand}, {Oman}, {Navarro}, \& {Simpson}}]{Cau20}
{Cautun}, M., {Ben{\'\i}tez-Llambay}, A., {Deason}, A.~J., {et~al.} 2020, \mnras, 494, 4291

\bibitem[{{Charbonnel} {et~al.}(2021){Charbonnel}, {Borisov}, {de Laverny}, \& {Prantzos}}]{Cha21}
{Charbonnel}, C., {Borisov}, S., {de Laverny}, P., \& {Prantzos}, N. 2021, \aap, 649, L10

\bibitem[{{Chavero} {et~al.}(2019){Chavero}, {de la Reza}, {Ghezzi}, {Llorente de Andr{\'e}s}, {Pereira}, {Giuppone}, \& {Pinz{\'o}n}}]{Cha19}
{Chavero}, C., {de la Reza}, R., {Ghezzi}, L., {et~al.} 2019, \mnras, 487, 3162

\bibitem[{{Chen} \& {Zhao}(2006)}]{Che06}
{Chen}, Y.~Q. \& {Zhao}, G. 2006, \aj, 131, 1816

\bibitem[{{Collier Cameron} \& {Jardine}(2018)}]{Col18}
{Collier Cameron}, A. \& {Jardine}, M. 2018, \mnras, 476, 2542

\bibitem[{{Dantas} {et~al.}(2022){Dantas}, {Guiglion}, {Smiljanic}, {Romano}, {Magrini}, {Bensby}, {Chiappini}, {Franciosini}, {Nepal}, {Tautvai{\v{s}}ien{\.{e}}}, {Gilmore}, {Randich}, {Lanzafame}, {Heiter}, {Morbidelli}, {Prisinzano}, \& {Zaggia}}]{Dan22}
{Dantas}, M.~L.~L., {Guiglion}, G., {Smiljanic}, R., {et~al.} 2022, \aap, 668, L7

\bibitem[{{Dantas} {et~al.}(2023){Dantas}, {Smiljanic}, {Boesso}, {Rocha-Pinto}, {Magrini}, {Guiglion}, {Tautvai{\v{s}}ien{\.{e}}}, {Gilmore}, {Randich}, {Bensby}, {Bragaglia}, {Bergemann}, {Carraro}, {Jofr{\'e}}, \& {Zaggia}}]{Dan23}
{Dantas}, M.~L.~L., {Smiljanic}, R., {Boesso}, R., {et~al.} 2023, \aap, 669, A96

\bibitem[{{de la Reza} {et~al.}(2023){de la Reza}, {Chavero}, {Roca-F{\`a}brega}, {Llorente de Andr{\'e}s}, {Cruz}, \& {Cifuentes}}]{Del23}
{de la Reza}, R., {Chavero}, C., {Roca-F{\`a}brega}, S., {et~al.} 2023, \aap, 671, A136

\bibitem[{{Deal} {et~al.}(2015){Deal}, {Richard}, \& {Vauclair}}]{Dea15}
{Deal}, M., {Richard}, O., \& {Vauclair}, S. 2015, \aap, 584, A105

\bibitem[{{Dehnen}(2000)}]{Deh00}
{Dehnen}, W. 2000, \aj, 119, 800

\bibitem[{{Delgado Mena} {et~al.}(2015){Delgado Mena}, {Bertr{\'a}n de Lis}, {Adibekyan}, {Sousa}, {Figueira}, {Mortier}, {Gonz{\'a}lez Hern{\'a}ndez}, {Tsantaki}, {Israelian}, \& {Santos}}]{Del15}
{Delgado Mena}, E., {Bertr{\'a}n de Lis}, S., {Adibekyan}, V.~Z., {et~al.} 2015, \aap, 576, A69

\bibitem[{{Delgado Mena} {et~al.}(2014){Delgado Mena}, {Israelian}, {Gonz{\'a}lez Hern{\'a}ndez}, {Sousa}, {Mortier}, {Santos}, {Adibekyan}, {Fernandes}, {Rebolo}, {Udry}, \& {Mayor}}]{Del14}
{Delgado Mena}, E., {Israelian}, G., {Gonz{\'a}lez Hern{\'a}ndez}, J.~I., {et~al.} 2014, \aap, 562, A92

\bibitem[{{Dumont} {et~al.}(2021){Dumont}, {Charbonnel}, {Palacios}, \& {Borisov}}]{Dum21}
{Dumont}, T., {Charbonnel}, C., {Palacios}, A., \& {Borisov}, S. 2021, \aap, 654, A46

\bibitem[{{Dumusque} {et~al.}(2011){Dumusque}, {Lovis}, {S{\'e}gransan}, {Mayor}, {Udry}, {Benz}, {Bouchy}, {Lo Curto}, {Mordasini}, {Pepe}, {Queloz}, {Santos}, \& {Naef}}]{Dum11}
{Dumusque}, X., {Lovis}, C., {S{\'e}gransan}, D., {et~al.} 2011, \aap, 535, A55

\bibitem[{{Eggenberger} {et~al.}(2012){Eggenberger}, {Haemmerl{\'e}}, {Meynet}, \& {Maeder}}]{Egg12}
{Eggenberger}, P., {Haemmerl{\'e}}, L., {Meynet}, G., \& {Maeder}, A. 2012, \aap, 539, A70

\bibitem[{{Franchini} {et~al.}(2014){Franchini}, {Morossi}, {di Marcantonio}, {Malagnini}, \& {Chavez}}]{Fra14}
{Franchini}, M., {Morossi}, C., {di Marcantonio}, P., {Malagnini}, M.~L., \& {Chavez}, M. 2014, \mnras, 442, 220

\bibitem[{{Fu} {et~al.}(2018){Fu}, {Romano}, {Bragaglia}, {Mucciarelli}, {Lind}, {Delgado Mena}, {Sousa}, {Randich}, {Bressan}, {Sbordone}, {Martell}, {Korn}, {Abia}, {Smiljanic}, {Jofr{\'e}}, {Pancino}, {Tautvai{\v{s}}ien{\.{e}}}, {Tang}, {Magrini}, {Lanzafame}, {Carraro}, {Bensby}, {Damiani}, {Alfaro}, {Flaccomio}, {Morbidelli}, {Zaggia}, {Lardo}, {Monaco}, {Frasca}, {Donati}, {Drazdauskas}, {Chorniy}, {Bayo}, \& {Kordopatis}}]{Fu2018}
{Fu}, X., {Romano}, D., {Bragaglia}, A., {et~al.} 2018, \aap, 610, A38

\bibitem[{{Gaia Collaboration} {et~al.}(2018){Gaia Collaboration}, {Brown}, {Vallenari}, {Prusti}, {de Bruijne}, {Babusiaux}, {Bailer-Jones}, {Biermann}, {Evans}, {Eyer}, {Jansen}, {Jordi}, {Klioner}, {Lammers}, {Lindegren}, {Luri}, {Mignard}, {Panem}, {Pourbaix}, {Randich}, {Sartoretti}, {Siddiqui}, {Soubiran}, {van Leeuwen}, {Walton}, {Arenou}, {Bastian}, {Cropper}, {Drimmel}, {Katz}, {Lattanzi}, {Bakker}, {Cacciari}, {Casta{\~n}eda}, {Chaoul}, {Cheek}, {De Angeli}, {Fabricius}, {Guerra}, {Holl}, {Masana}, {Messineo}, {Mowlavi}, {Nienartowicz}, {Panuzzo}, {Portell}, {Riello}, {Seabroke}, {Tanga}, {Th{\'e}venin}, {Gracia-Abril}, {Comoretto}, {Garcia-Reinaldos}, {Teyssier}, {Altmann}, {Andrae}, {Audard}, {Bellas-Velidis}, {Benson}, {Berthier}, {Blomme}, {Burgess}, {Busso}, {Carry}, {Cellino}, {Clementini}, {Clotet}, {Creevey}, {Davidson}, {De Ridder}, {Delchambre}, {Dell'Oro}, {Ducourant}, {Fern{\'a}ndez-Hern{\'a}ndez}, {Fouesneau}, {Fr{\'e}mat}, {Galluccio}, {Garc{\'\i}a-Torres},
  {Gonz{\'a}lez-N{\'u}{\~n}ez}, {Gonz{\'a}lez-Vidal}, {Gosset}, {Guy}, {Halbwachs}, {Hambly}, {Harrison}, {Hern{\'a}ndez}, {Hestroffer}, {Hodgkin}, {Hutton}, {Jasniewicz}, {Jean-Antoine-Piccolo}, {Jordan}, {Korn}, {Krone-Martins}, {Lanzafame}, {Lebzelter}, {L{\"o}ffler}, {Manteiga}, {Marrese}, {Mart{\'\i}n-Fleitas}, {Moitinho}, {Mora}, {Muinonen}, {Osinde}, {Pancino}, {Pauwels}, {Petit}, {Recio-Blanco}, {Richards}, {Rimoldini}, {Robin}, {Sarro}, {Siopis}, {Smith}, {Sozzetti}, {S{\"u}veges}, {Torra}, {van Reeven}, {Abbas}, {Abreu Aramburu}, {Accart}, {Aerts}, {Altavilla}, {{\'A}lvarez}, {Alvarez}, {Alves}, {Anderson}, {Andrei}, {Anglada Varela}, {Antiche}, {Antoja}, {Arcay}, {Astraatmadja}, {Bach}, {Baker}, {Balaguer-N{\'u}{\~n}ez}, {Balm}, {Barache}, {Barata}, {Barbato}, {Barblan}, {Barklem}, {Barrado}, {Barros}, {Barstow}, {Bartholom{\'e} Mu{\~n}oz}, {Bassilana}, {Becciani}, {Bellazzini}, {Berihuete}, {Bertone}, {Bianchi}, {Bienaym{\'e}}, {Blanco-Cuaresma}, {Boch}, {Boeche}, {Bombrun}, {Borrachero},
  {Bossini}, {Bouquillon}, {Bourda}, {Bragaglia}, {Bramante}, {Breddels}, {Bressan}, {Brouillet}, {Br{\"u}semeister}, {Brugaletta}, {Bucciarelli}, {Burlacu}, {Busonero}, {Butkevich}, {Buzzi}, {Caffau}, {Cancelliere}, {Cannizzaro}, {Cantat-Gaudin}, {Carballo}, {Carlucci}, {Carrasco}, {Casamiquela}, {Castellani}, {Castro-Ginard}, {Charlot}, {Chemin}, {Chiavassa}, {Cocozza}, {Costigan}, {Cowell}, {Crifo}, {Crosta}, {Crowley}, {Cuypers}, {Dafonte}, {Damerdji}, {Dapergolas}, {David}, {David}, {de Laverny}, {De Luise}, {De March}, {de Martino}, {de Souza}, {de Torres}, {Debosscher}, {del Pozo}, {Delbo}, {Delgado}, {Delgado}, {Di Matteo}, {Diakite}, {Diener}, {Distefano}, {Dolding}, {Drazinos}, {Dur{\'a}n}, {Edvardsson}, {Enke}, {Eriksson}, {Esquej}, {Eynard Bontemps}, {Fabre}, {Fabrizio}, {Faigler}, {Falc{\~a}o}, {Farr{\`a}s Casas}, {Federici}, {Fedorets}, {Fernique}, {Figueras}, {Filippi}, {Findeisen}, {Fonti}, {Fraile}, {Fraser}, {Fr{\'e}zouls}, {Gai}, {Galleti}, {Garabato}, {Garc{\'\i}a-Sedano}, {Garofalo},
  {Garralda}, {Gavel}, {Gavras}, {Gerssen}, {Geyer}, {Giacobbe}, {Gilmore}, {Girona}, {Giuffrida}, {Glass}, {Gomes}, {Granvik}, {Gueguen}, {Guerrier}, {Guiraud}, {Guti{\'e}rrez-S{\'a}nchez}, {Haigron}, {Hatzidimitriou}, {Hauser}, {Haywood}, {Heiter}, {Helmi}, {Heu}, {Hilger}, {Hobbs}, {Hofmann}, {Holland}, {Huckle}, {Hypki}, {Icardi}, {Jan{\ss}en}, {Jevardat de Fombelle}, {Jonker}, {Juh{\'a}sz}, {Julbe}, {Karampelas}, {Kewley}, {Klar}, {Kochoska}, {Kohley}, {Kolenberg}, {Kontizas}, {Kontizas}, {Koposov}, {Kordopatis}, {Kostrzewa-Rutkowska}, {Koubsky}, {Lambert}, {Lanza}, {Lasne}, {Lavigne}, {Le Fustec}, {Le Poncin-Lafitte}, {Lebreton}, {Leccia}, {Leclerc}, {Lecoeur-Taibi}, {Lenhardt}, {Leroux}, {Liao}, {Licata}, {Lindstr{\o}m}, {Lister}, {Livanou}, {Lobel}, {L{\'o}pez}, {Managau}, {Mann}, {Mantelet}, {Marchal}, {Marchant}, {Marconi}, {Marinoni}, {Marschalk{\'o}}, {Marshall}, {Martino}, {Marton}, {Mary}, {Massari}, {Matijevi{\v{c}}}, {Mazeh}, {McMillan}, {Messina}, {Michalik}, {Millar}, {Molina}, {Molinaro},
  {Moln{\'a}r}, {Montegriffo}, {Mor}, {Morbidelli}, {Morel}, {Morris}, {Mulone}, {Muraveva}, {Musella}, {Nelemans}, {Nicastro}, {Noval}, {O'Mullane}, {Ord{\'e}novic}, {Ord{\'o}{\~n}ez-Blanco}, {Osborne}, {Pagani}, {Pagano}, {Pailler}, {Palacin}, {Palaversa}, {Panahi}, {Pawlak}, {Piersimoni}, {Pineau}, {Plachy}, {Plum}, {Poggio}, {Poujoulet}, {Pr{\v{s}}a}, {Pulone}, {Racero}, {Ragaini}, {Rambaux}, {Ramos-Lerate}, {Regibo}, {Reyl{\'e}}, {Riclet}, {Ripepi}, {Riva}, {Rivard}, {Rixon}, {Roegiers}, {Roelens}, {Romero-G{\'o}mez}, {Rowell}, {Royer}, {Ruiz-Dern}, {Sadowski}, {Sagrist{\`a} Sell{\'e}s}, {Sahlmann}, {Salgado}, {Salguero}, {Sanna}, {Santana-Ros}, {Sarasso}, {Savietto}, {Schultheis}, {Sciacca}, {Segol}, {Segovia}, {S{\'e}gransan}, {Shih}, {Siltala}, {Silva}, {Smart}, {Smith}, {Solano}, {Solitro}, {Sordo}, {Soria Nieto}, {Souchay}, {Spagna}, {Spoto}, {Stampa}, {Steele}, {Steidelm{\"u}ller}, {Stephenson}, {Stoev}, {Suess}, {Surdej}, {Szabados}, {Szegedi-Elek}, {Tapiador}, {Taris}, {Tauran}, {Taylor},
  {Teixeira}, {Terrett}, {Teyssandier}, {Thuillot}, {Titarenko}, {Torra Clotet}, {Turon}, {Ulla}, {Utrilla}, {Uzzi}, {Vaillant}, {Valentini}, {Valette}, {van Elteren}, {Van Hemelryck}, {van Leeuwen}, {Vaschetto}, {Vecchiato}, {Veljanoski}, {Viala}, {Vicente}, {Vogt}, {von Essen}, {Voss}, {Votruba}, {Voutsinas}, {Walmsley}, {Weiler}, {Wertz}, {Wevers}, {Wyrzykowski}, {Yoldas}, {{\v{Z}}erjal}, {Ziaeepour}, {Zorec}, {Zschocke}, {Zucker}, {Zurbach}, \& {Zwitter}}]{GC18}
{Gaia Collaboration}, {Brown}, A.~G.~A., {Vallenari}, A., {et~al.} 2018, \aap, 616, A1

\bibitem[{{Gaia Collaboration} {et~al.}(2023){Gaia Collaboration}, {Vallenari}, {Brown}, {Prusti}, {de Bruijne}, {Arenou}, {Babusiaux}, {Biermann}, {Creevey}, {Ducourant}, {Evans}, {Eyer}, {Guerra}, {Hutton}, {Jordi}, {Klioner}, {Lammers}, {Lindegren}, {Luri}, {Mignard}, {Panem}, {Pourbaix}, {Randich}, {Sartoretti}, {Soubiran}, {Tanga}, {Walton}, {Bailer-Jones}, {Bastian}, {Drimmel}, {Jansen}, {Katz}, {Lattanzi}, {van Leeuwen}, {Bakker}, {Cacciari}, {Casta{\~n}eda}, {De Angeli}, {Fabricius}, {Fouesneau}, {Fr{\'e}mat}, {Galluccio}, {Guerrier}, {Heiter}, {Masana}, {Messineo}, {Mowlavi}, {Nicolas}, {Nienartowicz}, {Pailler}, {Panuzzo}, {Riclet}, {Roux}, {Seabroke}, {Sordo}, {Th{\'e}venin}, {Gracia-Abril}, {Portell}, {Teyssier}, {Altmann}, {Andrae}, {Audard}, {Bellas-Velidis}, {Benson}, {Berthier}, {Blomme}, {Burgess}, {Busonero}, {Busso}, {C{\'a}novas}, {Carry}, {Cellino}, {Cheek}, {Clementini}, {Damerdji}, {Davidson}, {de Teodoro}, {Nu{\~n}ez Campos}, {Delchambre}, {Dell'Oro}, {Esquej},
  {Fern{\'a}ndez-Hern{\'a}ndez}, {Fraile}, {Garabato}, {Garc{\'\i}a-Lario}, {Gosset}, {Haigron}, {Halbwachs}, {Hambly}, {Harrison}, {Hern{\'a}ndez}, {Hestroffer}, {Hodgkin}, {Holl}, {Jan{\ss}en}, {Jevardat de Fombelle}, {Jordan}, {Krone-Martins}, {Lanzafame}, {L{\"o}ffler}, {Marchal}, {Marrese}, {Moitinho}, {Muinonen}, {Osborne}, {Pancino}, {Pauwels}, {Recio-Blanco}, {Reyl{\'e}}, {Riello}, {Rimoldini}, {Roegiers}, {Rybizki}, {Sarro}, {Siopis}, {Smith}, {Sozzetti}, {Utrilla}, {van Leeuwen}, {Abbas}, {{\'A}brah{\'a}m}, {Abreu Aramburu}, {Aerts}, {Aguado}, {Ajaj}, {Aldea-Montero}, {Altavilla}, {{\'A}lvarez}, {Alves}, {Anders}, {Anderson}, {Anglada Varela}, {Antoja}, {Baines}, {Baker}, {Balaguer-N{\'u}{\~n}ez}, {Balbinot}, {Balog}, {Barache}, {Barbato}, {Barros}, {Barstow}, {Bartolom{\'e}}, {Bassilana}, {Bauchet}, {Becciani}, {Bellazzini}, {Berihuete}, {Bernet}, {Bertone}, {Bianchi}, {Binnenfeld}, {Blanco-Cuaresma}, {Blazere}, {Boch}, {Bombrun}, {Bossini}, {Bouquillon}, {Bragaglia}, {Bramante}, {Breedt},
  {Bressan}, {Brouillet}, {Brugaletta}, {Bucciarelli}, {Burlacu}, {Butkevich}, {Buzzi}, {Caffau}, {Cancelliere}, {Cantat-Gaudin}, {Carballo}, {Carlucci}, {Carnerero}, {Carrasco}, {Casamiquela}, {Castellani}, {Castro-Ginard}, {Chaoul}, {Charlot}, {Chemin}, {Chiaramida}, {Chiavassa}, {Chornay}, {Comoretto}, {Contursi}, {Cooper}, {Cornez}, {Cowell}, {Crifo}, {Cropper}, {Crosta}, {Crowley}, {Dafonte}, {Dapergolas}, {David}, {David}, {de Laverny}, {De Luise}, {De March}, {De Ridder}, {de Souza}, {de Torres}, {del Peloso}, {del Pozo}, {Delbo}, {Delgado}, {Delisle}, {Demouchy}, {Dharmawardena}, {Di Matteo}, {Diakite}, {Diener}, {Distefano}, {Dolding}, {Edvardsson}, {Enke}, {Fabre}, {Fabrizio}, {Faigler}, {Fedorets}, {Fernique}, {Fienga}, {Figueras}, {Fournier}, {Fouron}, {Fragkoudi}, {Gai}, {Garcia-Gutierrez}, {Garcia-Reinaldos}, {Garc{\'\i}a-Torres}, {Garofalo}, {Gavel}, {Gavras}, {Gerlach}, {Geyer}, {Giacobbe}, {Gilmore}, {Girona}, {Giuffrida}, {Gomel}, {Gomez}, {Gonz{\'a}lez-N{\'u}{\~n}ez},
  {Gonz{\'a}lez-Santamar{\'\i}a}, {Gonz{\'a}lez-Vidal}, {Granvik}, {Guillout}, {Guiraud}, {Guti{\'e}rrez-S{\'a}nchez}, {Guy}, {Hatzidimitriou}, {Hauser}, {Haywood}, {Helmer}, {Helmi}, {Sarmiento}, {Hidalgo}, {Hilger}, {H{\l}adczuk}, {Hobbs}, {Holland}, {Huckle}, {Jardine}, {Jasniewicz}, {Jean-Antoine Piccolo}, {Jim{\'e}nez-Arranz}, {Jorissen}, {Juaristi Campillo}, {Julbe}, {Karbevska}, {Kervella}, {Khanna}, {Kontizas}, {Kordopatis}, {Korn}, {K{\'o}sp{\'a}l}, {Kostrzewa-Rutkowska}, {Kruszy{\'n}ska}, {Kun}, {Laizeau}, {Lambert}, {Lanza}, {Lasne}, {Le Campion}, {Lebreton}, {Lebzelter}, {Leccia}, {Leclerc}, {Lecoeur-Taibi}, {Liao}, {Licata}, {Lindstr{\o}m}, {Lister}, {Livanou}, {Lobel}, {Lorca}, {Loup}, {Madrero Pardo}, {Magdaleno Romeo}, {Managau}, {Mann}, {Manteiga}, {Marchant}, {Marconi}, {Marcos}, {Marcos Santos}, {Mar{\'\i}n Pina}, {Marinoni}, {Marocco}, {Marshall}, {Martin Polo}, {Mart{\'\i}n-Fleitas}, {Marton}, {Mary}, {Masip}, {Massari}, {Mastrobuono-Battisti}, {Mazeh}, {McMillan}, {Messina}, {Michalik},
  {Millar}, {Mints}, {Molina}, {Molinaro}, {Moln{\'a}r}, {Monari}, {Mongui{\'o}}, {Montegriffo}, {Montero}, {Mor}, {Mora}, {Morbidelli}, {Morel}, {Morris}, {Muraveva}, {Murphy}, {Musella}, {Nagy}, {Noval}, {Oca{\~n}a}, {Ogden}, {Ordenovic}, {Osinde}, {Pagani}, {Pagano}, {Palaversa}, {Palicio}, {Pallas-Quintela}, {Panahi}, {Payne-Wardenaar}, {Pe{\~n}alosa Esteller}, {Penttil{\"a}}, {Pichon}, {Piersimoni}, {Pineau}, {Plachy}, {Plum}, {Poggio}, {Pr{\v{s}}a}, {Pulone}, {Racero}, {Ragaini}, {Rainer}, {Raiteri}, {Rambaux}, {Ramos}, {Ramos-Lerate}, {Re Fiorentin}, {Regibo}, {Richards}, {Rios Diaz}, {Ripepi}, {Riva}, {Rix}, {Rixon}, {Robichon}, {Robin}, {Robin}, {Roelens}, {Rogues}, {Rohrbasser}, {Romero-G{\'o}mez}, {Rowell}, {Royer}, {Ruz Mieres}, {Rybicki}, {Sadowski}, {S{\'a}ez N{\'u}{\~n}ez}, {Sagrist{\`a} Sell{\'e}s}, {Sahlmann}, {Salguero}, {Samaras}, {Sanchez Gimenez}, {Sanna}, {Santove{\~n}a}, {Sarasso}, {Schultheis}, {Sciacca}, {Segol}, {Segovia}, {S{\'e}gransan}, {Semeux}, {Shahaf}, {Siddiqui}, {Siebert},
  {Siltala}, {Silvelo}, {Slezak}, {Slezak}, {Smart}, {Snaith}, {Solano}, {Solitro}, {Souami}, {Souchay}, {Spagna}, {Spina}, {Spoto}, {Steele}, {Steidelm{\"u}ller}, {Stephenson}, {S{\"u}veges}, {Surdej}, {Szabados}, {Szegedi-Elek}, {Taris}, {Taylor}, {Teixeira}, {Tolomei}, {Tonello}, {Torra}, {Torra}, {Torralba Elipe}, {Trabucchi}, {Tsounis}, {Turon}, {Ulla}, {Unger}, {Vaillant}, {van Dillen}, {van Reeven}, {Vanel}, {Vecchiato}, {Viala}, {Vicente}, {Voutsinas}, {Weiler}, {Wevers}, {Wyrzykowski}, {Yoldas}, {Yvard}, {Zhao}, {Zorec}, {Zucker}, \& {Zwitter}}]{gaiaDR3}
{Gaia Collaboration}, {Vallenari}, A., {Brown}, A.~G.~A., {et~al.} 2023, \aap, 674, A1

\bibitem[{{Ghezzi} {et~al.}(2010){Ghezzi}, {Cunha}, {Smith}, \& {de la Reza}}]{Ghe10}
{Ghezzi}, L., {Cunha}, K., {Smith}, V.~V., \& {de la Reza}, R. 2010, \apj, 724, 154

\bibitem[{{Gonzalez}(2014)}]{Gon14}
{Gonzalez}, G. 2014, \mnras, 441, 1201

\bibitem[{{Gonzalez} {et~al.}(2010){Gonzalez}, {Carlson}, \& {Tobin}}]{Gon10}
{Gonzalez}, G., {Carlson}, M.~K., \& {Tobin}, R.~W. 2010, \mnras, 403, 1368

\bibitem[{{Gonzalez} \& {Laws}(2000)}]{Gon00}
{Gonzalez}, G. \& {Laws}, C. 2000, \aj, 119, 390

\bibitem[{{Guiglion} {et~al.}(2019){Guiglion}, {Chiappini}, {Romano}, {Matteucci}, {Anders}, {Steinmetz}, {Minchev}, {de Laverny}, \& {Recio-Blanco}}]{Gui19}
{Guiglion}, G., {Chiappini}, C., {Romano}, D., {et~al.} 2019, \aap, 623, A99

\bibitem[{{Guiglion} {et~al.}(2016){Guiglion}, {de Laverny}, {Recio-Blanco}, {Worley}, {De Pascale}, {Masseron}, {Prantzos}, \& {Mikolaitis}}]{Gui16}
{Guiglion}, G., {de Laverny}, P., {Recio-Blanco}, A., {et~al.} 2016, \aap, 595, A18

\bibitem[{{Hamer} \& {Schlaufman}(2019)}]{Ham19}
{Hamer}, J.~H. \& {Schlaufman}, K.~C. 2019, \aj, 158, 190

\bibitem[{{Isaacson} \& {Fischer}(2010)}]{Isa10}
{Isaacson}, H. \& {Fischer}, D. 2010, \apj, 725, 875

\bibitem[{{Israelian} {et~al.}(2004){Israelian}, {Santos}, {Mayor}, \& {Rebolo}}]{Isr04}
{Israelian}, G., {Santos}, N.~C., {Mayor}, M., \& {Rebolo}, R. 2004, \aap, 414, 601

\bibitem[{{Kervella} {et~al.}(2019){Kervella}, {Arenou}, {Mignard}, \& {Th{\'e}venin}}]{Ke19}
{Kervella}, P., {Arenou}, F., {Mignard}, F., \& {Th{\'e}venin}, F. 2019, \aap, 623, A72

\bibitem[{{Lai}(2012)}]{Lai12}
{Lai}, D. 2012, \mnras, 423, 486

\bibitem[{{Lazovik}(2021)}]{Laz21}
{Lazovik}, Y.~A. 2021, \mnras, 508, 3408

\bibitem[{{Leconte} {et~al.}(2010){Leconte}, {Chabrier}, {Baraffe}, \& {Levrard}}]{Lec10}
{Leconte}, J., {Chabrier}, G., {Baraffe}, I., \& {Levrard}, B. 2010, \aap, 516, A64

\bibitem[{{Lee-Brown}(2018)}]{Lee18}
{Lee-Brown}, D.~B. 2018, PhD thesis, University of Kansas

\bibitem[{{Lineweaver} \& {Grether}(2003)}]{Lin03}
{Lineweaver}, C.~H. \& {Grether}, D. 2003, \apj, 598, 1350

\bibitem[{{Llorente de Andr{\'e}s} {et~al.}(2021){Llorente de Andr{\'e}s}, {Chavero}, {de la Reza}, {Roca-F{\`a}brega}, \& {Cifuentes}}]{Llo21}
{Llorente de Andr{\'e}s}, F., {Chavero}, C., {de la Reza}, R., {Roca-F{\`a}brega}, S., \& {Cifuentes}, C. 2021, \aap, 654, A137

\bibitem[{{Llorente de Andr{\'e}s} \& {Morales-Dur{\'a}n}(2022)}]{Llo22}
{Llorente de Andr{\'e}s}, F. \& {Morales-Dur{\'a}n}, C. 2022, American Journal of Astronomy and Astrophysiscs, 9, 52

\bibitem[{{Luck} \& {Heiter}(2006)}]{Luc06}
{Luck}, R.~E. \& {Heiter}, U. 2006, \aj, 131, 3069

\bibitem[{{Ma} \& {Fuller}(2021)}]{Ma21}
{Ma}, L. \& {Fuller}, J. 2021, \apj, 918, 16

\bibitem[{{Marsden} {et~al.}(2014){Marsden}, {Petit}, {Jeffers}, {Morin}, {Fares}, {Reiners}, {do Nascimento}, {Auri{\`e}re}, {Bouvier}, {Carter}, {Catala}, {Dintrans}, {Donati}, {Gastine}, {Jardine}, {Konstantinova-Antova}, {Lanoux}, {Ligni{\`e}res}, {Morgenthaler}, {Ram{\`\i}rez-V{\`e}lez}, {Th{\'e}ado}, {Van Grootel}, \& {BCool Collaboration}}]{Mar14}
{Marsden}, S.~C., {Petit}, P., {Jeffers}, S.~V., {et~al.} 2014, \mnras, 444, 3517

\bibitem[{{Martos} {et~al.}(2023){Martos}, {Mel{\'e}ndez}, {Rathsam}, \& {Carvalho Silva}}]{Mar23}
{Martos}, G., {Mel{\'e}ndez}, J., {Rathsam}, A., \& {Carvalho Silva}, G. 2023, \mnras, 522, 3217

\bibitem[{{Matsumura} {et~al.}(2010){Matsumura}, {Peale}, \& {Rasio}}]{Mat10}
{Matsumura}, S., {Peale}, S.~J., \& {Rasio}, F.~A. 2010, \apj, 725, 1995

\bibitem[{{Mel{\'e}ndez}(2020)}]{Mel20}
{Mel{\'e}ndez}, J. 2020, Astronomische Nachrichten, 341, 493

\bibitem[{{Metzger} {et~al.}(2012){Metzger}, {Giannios}, \& {Spiegel}}]{Met12}
{Metzger}, B.~D., {Giannios}, D., \& {Spiegel}, D.~S. 2012, \mnras, 425, 2778

\bibitem[{{Mishenina} {et~al.}(2012){Mishenina}, {Soubiran}, {Kovtyukh}, {Katsova}, \& {Livshits}}]{Mis2012}
{Mishenina}, T.~V., {Soubiran}, C., {Kovtyukh}, V.~V., {Katsova}, M.~M., \& {Livshits}, M.~A. 2012, \aap, 547, A106

\bibitem[{{Mulders}(2018)}]{Mul18}
{Mulders}, G.~D. 2018, in Handbook of Exoplanets, ed. H.~J. {Deeg} \& J.~A. {Belmonte}, 153

\bibitem[{{Munoz Romero} \& {Kempton}(2018)}]{Mun18}
{Munoz Romero}, C.~E. \& {Kempton}, E. M.~R. 2018, \aj, 155, 134

\bibitem[{{Oelkers} {et~al.}(2018){Oelkers}, {Rodriguez}, {Stassun}, {Pepper}, {Somers}, {Kafka}, {Stevens}, {Beatty}, {Siverd}, {Lund}, {Kuhn}, {James}, \& {Gaudi}}]{Oel18}
{Oelkers}, R.~J., {Rodriguez}, J.~E., {Stassun}, K.~G., {et~al.} 2018, \aj, 155, 39

\bibitem[{{Ogilvie}(2014)}]{Ogi14}
{Ogilvie}, G.~I. 2014, \araa, 52, 171

\bibitem[{{Oh} {et~al.}(2018){Oh}, {Price-Whelan}, {Brewer}, {Hogg}, {Spergel}, \& {Myles}}]{Oh2018}
{Oh}, S., {Price-Whelan}, A.~M., {Brewer}, J.~M., {et~al.} 2018, \apj, 854, 138

\bibitem[{{P{\~o}der} {et~al.}(2023){P{\~o}der}, {Benito}, {Pata}, {Kipper}, {Ramler}, {H{\"u}tsi}, {Kolka}, \& {Thomas}}]{Pod23}
{P{\~o}der}, S., {Benito}, M., {Pata}, J., {et~al.} 2023, \aap, 676, A134

\bibitem[{{Pasquini} {et~al.}(2007){Pasquini}, {D{\"o}llinger}, {Weiss}, {Girardi}, {Chavero}, {Hatzes}, {da Silva}, \& {Setiawan}}]{Pas07}
{Pasquini}, L., {D{\"o}llinger}, M.~P., {Weiss}, A., {et~al.} 2007, \aap, 473, 979

\bibitem[{{Paxton} {et~al.}(2011){Paxton}, {Bildsten}, {Dotter}, {Herwig}, {Lesaffre}, \& {Timmes}}]{Pax11}
{Paxton}, B., {Bildsten}, L., {Dotter}, A., {et~al.} 2011, \apjs, 192, 3

\bibitem[{{Paxton} {et~al.}(2019){Paxton}, {Smolec}, {Schwab}, {Gautschy}, {Bildsten}, {Cantiello}, {Dotter}, {Farmer}, {Goldberg}, {Jermyn}, {Kanbur}, {Marchant}, {Thoul}, {Townsend}, {Wolf}, {Zhang}, \& {Timmes}}]{Pax19}
{Paxton}, B., {Smolec}, R., {Schwab}, J., {et~al.} 2019, \apjs, 243, 10

\bibitem[{{Penev} {et~al.}(2018){Penev}, {Bouma}, {Winn}, \& {Hartman}}]{Pen18}
{Penev}, K., {Bouma}, L.~G., {Winn}, J.~N., \& {Hartman}, J.~D. 2018, \aj, 155, 165

\bibitem[{{Perdigon} {et~al.}(2021){Perdigon}, {de Laverny}, {Recio-Blanco}, {Fernandez-Alvar}, {Santos-Peral}, {Kordopatis}, \& {{\'A}lvarez}}]{Per21}
{Perdigon}, J., {de Laverny}, P., {Recio-Blanco}, A., {et~al.} 2021, \aap, 647, A162

\bibitem[{{Perryman}(2018)}]{Per18}
{Perryman}, M. 2018, {The Exoplanet Handbook}

\bibitem[{{Ram{\'\i}rez} {et~al.}(2012){Ram{\'\i}rez}, {Fish}, {Lambert}, \& {Allende Prieto}}]{Ram12}
{Ram{\'\i}rez}, I., {Fish}, J.~R., {Lambert}, D.~L., \& {Allende Prieto}, C. 2012, \apj, 756, 46

\bibitem[{{Randich} {et~al.}(2020){Randich}, {Pasquini}, {Franciosini}, {Magrini}, {Jackson}, {Jeffries}, {d'Orazi}, {Romano}, {Sanna}, {Tautvai{\v{s}}ien{\.{e}}}, {Tsantaki}, {Wright}, {Gilmore}, {Bensby}, {Bragaglia}, {Pancino}, {Smiljanic}, {Bayo}, {Carraro}, {Gonneau}, {Hourihane}, {Morbidelli}, \& {Worley}}]{Ran20}
{Randich}, S., {Pasquini}, L., {Franciosini}, E., {et~al.} 2020, \aap, 640, L1

\bibitem[{{Rice} \& {Brewer}(2020)}]{Ric20}
{Rice}, M. \& {Brewer}, J.~M. 2020, \apj, 898, 119

\bibitem[{{Roca-F{\`a}brega} {et~al.}(2021){Roca-F{\`a}brega}, {Llorente de Andr{\'e}s}, {Chavero}, {Cifuentes}, \& {de la Reza}}]{Roc21}
{Roca-F{\`a}brega}, S., {Llorente de Andr{\'e}s}, F., {Chavero}, C., {Cifuentes}, C., \& {de la Reza}, R. 2021, \aap, 656, A64

\bibitem[{{Romano} {et~al.}(2021){Romano}, {Magrini}, {Randich}, {Casali}, {Bonifacio}, {Jeffries}, {Matteucci}, {Franciosini}, {Spina}, {Guiglion}, {Chiappini}, {Mucciarelli}, {Ventura}, {Grisoni}, {Bellazzini}, {Bensby}, {Bragaglia}, {de Laverny}, {Korn}, {Martell}, {Tautvai{\v{s}}ien{\.{e}}}, {Carraro}, {Gonneau}, {Jofr{\'e}}, {Pancino}, {Smiljanic}, {Vallenari}, {Fu}, {Guti{\'e}rrez Albarr{\'a}n}, {Jim{\'e}nez-Esteban}, {Montes}, {Damiani}, {Bergemann}, \& {Worley}}]{Rom21}
{Romano}, D., {Magrini}, L., {Randich}, S., {et~al.} 2021, \aap, 653, A72

\bibitem[{{Ryan}(2000)}]{Rya00}
{Ryan}, S.~G. 2000, \mnras, 316, L35

\bibitem[{{Sevilla} {et~al.}(2022){Sevilla}, {Behmard}, \& {Fuller}}]{Sev22}
{Sevilla}, J., {Behmard}, A., \& {Fuller}, J. 2022, \mnras, 516, 3354

\bibitem[{{Soares-Furtado} {et~al.}(2021){Soares-Furtado}, {Cantiello}, {MacLeod}, \& {Ness}}]{Soa21}
{Soares-Furtado}, M., {Cantiello}, M., {MacLeod}, M., \& {Ness}, M.~K. 2021, \aj, 162, 273

\bibitem[{{Soubiran} {et~al.}(2016){Soubiran}, {Le Campion}, {Brouillet}, \& {Chemin}}]{Sou16}
{Soubiran}, C., {Le Campion}, J.-F., {Brouillet}, N., \& {Chemin}, L. 2016, \aap, 591, A118

\bibitem[{{Spiegel} {et~al.}(2011){Spiegel}, {Burrows}, \& {Milsom}}]{Spi11}
{Spiegel}, D.~S., {Burrows}, A., \& {Milsom}, J.~A. 2011, \apj, 727, 57

\bibitem[{{Stephan} {et~al.}(2020){Stephan}, {Naoz}, {Gaudi}, \& {Salas}}]{Ste20}
{Stephan}, A.~P., {Naoz}, S., {Gaudi}, B.~S., \& {Salas}, J.~M. 2020, \apj, 889, 45

\bibitem[{{Sun} {et~al.}(2008){Sun}, {Ferraz-Mello}, \& {Zhou}}]{Sun08}
{Sun}, Y.-S., {Ferraz-Mello}, S., \& {Zhou}, J.-L. 2008, {Exoplanets: Detection, Formation and Dynamics (IAU S249)}

\bibitem[{{Takeda} {et~al.}(2010){Takeda}, {Honda}, {Kawanomoto}, {Ando}, \& {Sakurai}}]{Tak10}
{Takeda}, Y., {Honda}, S., {Kawanomoto}, S., {Ando}, H., \& {Sakurai}, T. 2010, \aap, 515, A93

\bibitem[{{Takeda} \& {Kawanomoto}(2005)}]{Tak05}
{Takeda}, Y. \& {Kawanomoto}, S. 2005, \pasj, 57, 45

\bibitem[{{Th{\'e}ado} \& {Vauclair}(2012)}]{The12}
{Th{\'e}ado}, S. \& {Vauclair}, S. 2012, \apj, 744, 123

\bibitem[{{Wright} {et~al.}(2012){Wright}, {Marcy}, {Howard}, {Johnson}, {Morton}, \& {Fischer}}]{Wri12}
{Wright}, J.~T., {Marcy}, G.~W., {Howard}, A.~W., {et~al.} 2012, \apj, 753, 160

\end{thebibliography}


\begin{appendix}
    
\section{Adopted stellar parameters}\label{appenA}

\begin{table*}[!ht]
\caption{Parameters of stars and their host planets, complementary to the sample published in LA21.}
\label{t:parameters0}
\centering
\begin{tabular}{lccccccccccc}
\hline
\hline   
   Identifier & name & Spectral & Nbr. & $\Sigma M_p$ & P & a & ecc & Prot & Rotation & $\log{\rm OD}$ & Radii\\
   & & type & Planet & (M$_{\rm Jup}$) & (days) & (au) &  & (days) & ref. & & ref. \\ 
    \hline
HD\,142	&	HD142 A	&	F7V	&	2	&	6.55	&	349.70	&	1.02	&	0.17	&	27.15	&	Oe18	&	6.757	&	Ke19	\\
HD\,1461	&	HD 1461  	&	G5V	&	2	&	0.04	&	5.77	&	0.06	&	0.00	&	29.00	&	Ma14	&	1.608	&	Ke19	\\
HD\,2039	&	HD 2039  	&	G2/3IV/V	&	1	&	6.11	&	1120.00	&	2.23	&	0.67	&	\ldots	&	\ldots	&	9.284	&	Ke19	\\
HD\,6434	&	HD 6434  	&	G2/3V	&	1	&	0.39	&	22.00	&	0.14	&	0.17	&	\ldots	&	\ldots	&	\ldots	&	Ke19	\\
HD\,6718	&	HD 6718  	&	G5V	&	1	&	1.56	&	2496.00	&	3.56	&	0.10	&	\ldots	&	\ldots	&	11.550	&	Ke19	\\
HD\,7449	&	HD 7449  	&	G0V	&	2	&	3.11	&	1275.00	&	2.30	&	0.50	&	13.30	&	Du11	&	10.061	&	Ke19	\\
HD\,8535	&	HD 8535  	&	G0V	&	1	&	0.68	&	1313.00	&	2.45	&	0.15	&	26.44	&	Oe18	&	10.283	&	Ke19	\\
HD\,8574	&	HD 8574  	&	F8	&	1	&	2.11	&	227.55	&	0.77	&	0.29	&	15.00	&	Is10	&	6.400	&	Ke19	\\
HD\,9578	&	HD 9578  	&	G1V	&	1	&	0.62	&	494.00	&	…	&	0.00	&	\ldots	&	\ldots	&	8.833	&	Ke19	\\
HD\,9446	&	HD 9446  	&	G5V	&	2	&	2.52	&	30.05	&	0.19	&	0.06	&	0.88	&	Oe18	&	3.176	&	Ke19	\\
...	&	...  	&	...	&	...	&	...	&	...	&	...	&	...	&	...	&	...	&	...	&	...	\\
  \hline

\end{tabular}
\tablefoot{This table is only partially presented. Its complete version is available online, as a Vizier catalogue.\\
{\bf Ref.} Du11 - \citet{Dum11}; Is10 - \citet{Isa10}; Ke19 - \citet{Ke19}; Ma14 - \citet{Mar14}; Oe18 - \citet{Oel18}.}

\end{table*}   

\begin{table}
\caption{[Ti/H] and [Ti/Fe] values for objects with negative $U$ that are older than 8\,Ga.} 
\label{tableA2}
\centering
\begin{tabular}{lccc}
\hline
\hline   
   Identifier & [Ti/H]  & [Ti/Fe] & Reference \\
   & (dex) &  (dex) & \\ 
    \hline

CD$-$45\,12460	&	$-$0.66	&	0.2	&	Be14		\\
HD\,102158	&	$-$0.133	&	0.327	&	Ri20		\\
HD\,102200	&	$-$1.05	&	0.06	&	Be14		\\
HD\,102300	&	0.11	&	0.11$^*$	&	Pe21	\\
HD\,104800	&	$-$0.56	&	0.24	&	Be14		\\
HD\,108204	&	0.26	&	0.26$^*$	&	Fra14	\\
HD\,108309	&	0.08	&	$-$0.04	&	Be14		\\
HD\,109684	&	0.09	&	0.09$^*$	&	Pe21	\\
HD\,112257	&	0.044	&	0.074	&	Ri20		\\
HD\,113679	&	$-$0.34	&	0.27	&	Be14		\\
...	&	...	&	...	&	...		\\
 
 \hline 
  \hline
  
 \end{tabular}                                                                

\tablefoot{ The [Ti/Fe] values marked with an asterisk are [$\alpha$/Fe], as the latter is adopted when [Ti/Fe] is not available. This table is only partially presented. Its complete version is available online, as a Vizier catalogue.\\
{\bf Ref.} Be14 - \citet{Ben14}; Fra14 - \citep{Fra14}; Pe21 - \citet{Per21}; Ri20 - \citet[][]{Ric20}.}

\end{table}  

Table~\ref{t:parameters0} shows the parameters of stars with detected planets, as described in Sect.~\ref{sec:sample}. This table is complementary to the table published in \citet{Llo21}. The last column presents the calculated orbital decay (as $\log{\rm OD}$), as described Sect.~\ref{sec:interaction}. 

Table~\ref{tableA2} shows [Ti/H] and [Ti/Fe] values for the objects with a negative Galactic velocity, $U$, that are older than 8\,Ga, as described in Sect.~\ref{modonmi}. Values marked with an asterisk are [a/Fe] instead, when [Ti/Fe] is not available.
 
\end{appendix}

\end{document}